# Universal Thickness-Dependent Absorption in Solids at the Nanoscale: Anomalous Enhancement in the Ultrathin Limit

*Bhumika Chauhan[1], Nikhil Singh[1], Subhrajit Dalai[1], Abhisek Saidarsan[1], Sayantan Patra[1], Sourabh Jain[1], Aparna Deshpande[1], Ashish Arora[1,*]*

[1]*Department of Physics, Indian Institute of Science Education and Research, Dr. Homi Bhabha Road, 411008 Pune, India*

*Email : ashish.arora@iiserpune.ac.in

Through systematic experimental and theoretical studies of layer-thickness-dependent absorption in semiconducting $MoSe_2$ and $WS_2$ across the visible to near-infrared spectral range, we demonstrate a universal absorption behavior in solids at nanoscale thicknesses. With increasing thickness, a non-monotonic evolution of absorption integrated over the measured spectral region is revealed which is accompanied by pronounced oscillatory features. This shows a strong deviation from the expected Beer-Lambert law. Below 10 nm, we observe a sharp anomalous increase in absorption, with deviations from Beer's law exceeding 50% in layered semiconductors. Our conclusions hold irrespective of the presence of any optical resonances such as excitons or plasmons within the spectral window. The observed behavior has origins in the electromagnetic interference effects taking place between the two surfaces of the thin crystals. The present work on 2D semiconductors is extendable to all kinds of solids such as conventional semiconductors (e.g. Si, GaAs, GaN, InP), (semi) metals (e.g. Al, Ag, Au, c-HOPG) and 2D magnetic materials (e.g. CrSBr and $NiPS_3$). Our results provide fundamental insights into light-matter interactions in solids at the nanoscale and are vital for optimally designing the new generation of absorption-based flexible optoelectronic devices.

Low-dimensional solids such as quantum wells and van der Waals (vdW) materials are excellent platforms to study two-dimensional quantum effects and for their potential towards the next generation of optoelectronic, spintronic and valleytronic devices [1–5]. Group III-V and II-VI semiconductor quantum wells and heterostructures have revolutionized modern optoelectronics [6]. Recently, vdW materials such as transition metal dichalcogenides (TMDCs) of chemical formula $MX_2$ ($M$ = Mo, W, Re etc., $X$ = S, Se, Te etc.), graphene, hexagonal boron nitride (hBN), and the newly emerging 2D magnetic materials (2DMM e.g. $CrX_3$ with $X$ = Cl, Br, I; $MPX_3$ with $M$ = Mn, Fe, Zn, Ni and $X$ = S, Se; $Fe_nGeTe_2$ with n = 3,4 and 5) have opened immense potential towards flexible and energy-efficient device technologies [3,7]. Semiconductors are active layers in the absorption-based optoelectronic devices such as photodiodes, photovoltaics, phototransistors, avalanche photodiodes, charge coupled devices (CCDs) and complementary metal-oxide-semiconductor (CMOS) devices [6]. Interestingly, monolayer TMDCs, despite their ultrathin thickness (less than 1nm), can absorb a considerable fraction of the visible light (approximately 5-10 %) [8]. Additionally, spintronics and valleytronics require heterostructures of thin semiconductor films (such as TMDCs) with magnetic films (such as 2DMMs) for efficient control of spin-valley degrees of freedom in TMDCs [4]. Indeed, a complete device also requires electrical contacts through thin metallic/semi-metallic films such as silver, gold or graphene [7,9]. Overall, it is crucial to understand thickness-dependent light absorption by thin solid films for an optimal device performance. This is especially important for photovoltaic devices where optimized film thicknesses are essential for large efficiencies [9–11]. Unfortunately, a careful study of absorption of thin solid films with layer thickness is not available to the best of our knowledge [12]. As such, before layered materials era, it was challenging for researchers to prepare samples suited to measure absorption with varying layer thicknesses.

Intuition, and experimental studies on some solids such as silicon covered with silica nanoparticles [13], $V_2O_5$ [14], $SiO_2$ [15], $MgF_2$ [15] or TiN [16] suggest that the absorption of light in a material should monotonically increase with increasing layer thickness, and should follow the well-established Beer-Lambert law [12,17–19]. However, for layer thicknesses in the sub-wavelength regime, which are required for the ultrathin device applications, strong interference effects can lead to drastic deviations from intuitive expectations [20,21]. Some well-known interference-induced effects include oscillatory reflectance contrast of layered materials on $SiO_2$/Si substrates as a function of $SiO_2$ thickness [22], or reflectance/transmittance as a function of wavelength, or thickness at a certain wavelength [15,23,24]. One previous work discovered an oscillatory absorption at the $A$ exciton line in $WSe_2$ and $MoSe_2$ crystals as a function of layer thickness [25]. The authors explained their observations on the basis of super-radiant coupling of excitonic emitters via creation of long-lived exciton polaritons.

In the present work, we systematically address this issue of thickness-dependent absorption both theoretically and experimentally. We measure absorption spectra of thin crystals of $MoSe_2$ and $WS_2$ as representative materials, with layer thicknesses up to 200 nm. We observe striking deviations from the Beer–Lambert law, with discrepancies exceeding 50% in these layered semiconductors at thicknesses below 10 nm. This phenomenon is explained using a model based on the generalized transfer-matrix (GTM) method, which addresses interference effects taking place between multiple interfaces of the samples. We see no evidence of super-radiant coupling of excitonic emitters as reported earlier [25]. We extend our results theoretically to various types of solids such as conventional semiconductors (e.g. GaAs, Si, InP, GaN), 2DMMs (e.g. CrSBr, $NiPS_3$) and (semi)metals (e.g. Au, Al, Ag, graphite) and find universal trends.

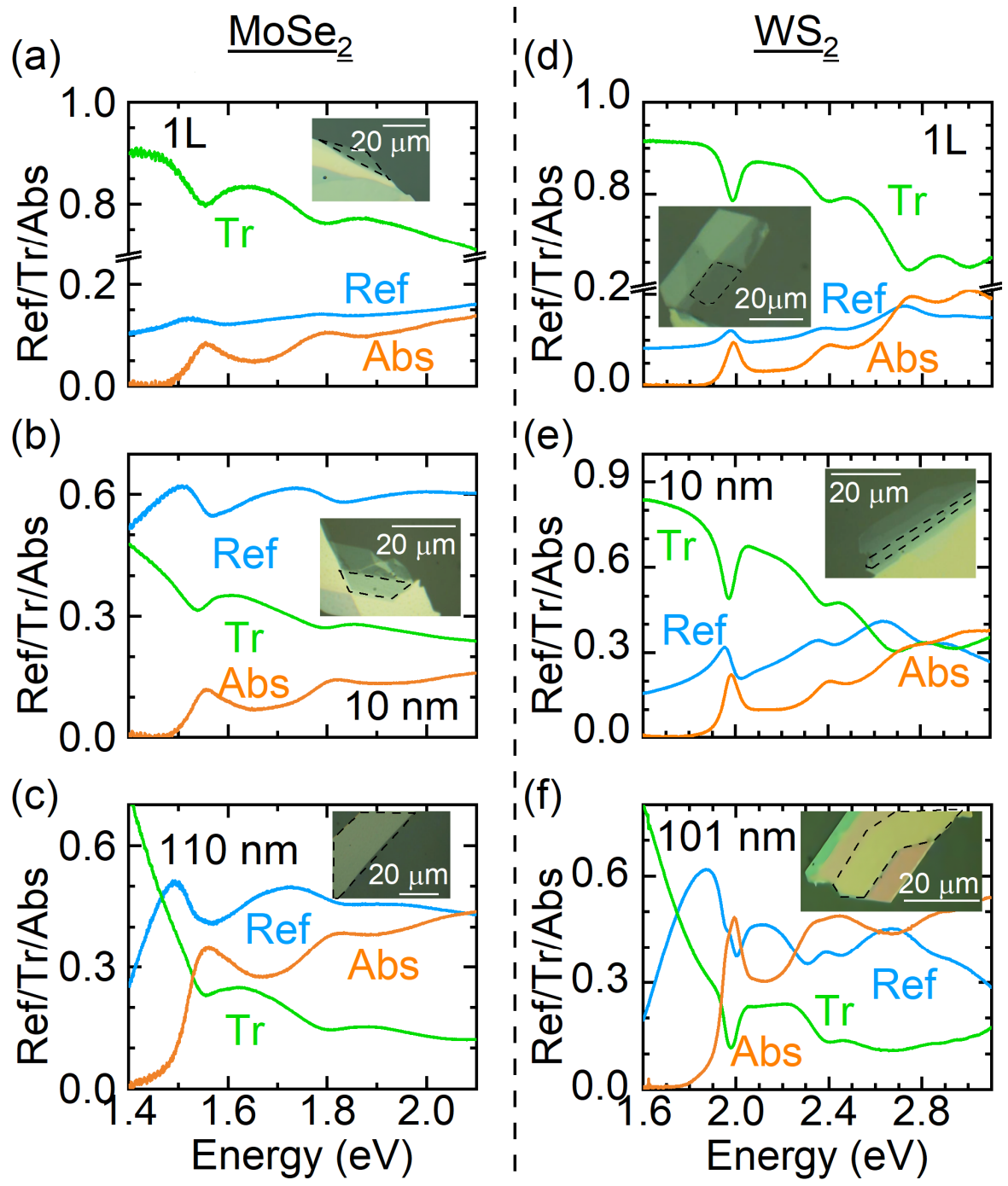

Figure 1. Reflectance (Ref, blue), transmittance (Tr, green), and absorption (Abs, orange) spectra of monolayer (1L), 10 nm, and 110 nm $MoSe_2$ (left), and 1L, 10 nm and 101 nm thick $WS_2$ (right). The crystals are exfoliated on 0.5 mm thick sapphire substrate. Insets show optical microscope images of the crystals, where regions enclosed by dashed lines are the flakes for which data are presented.

The samples used for our experiments are prepared using the standard exfoliation and viscoelastic dry-transfer method [26]. We exfoliate thin layers of $MoSe_2$ and $WS_2$ single crystals (obtained from 2D semiconductors, USA) on 0.5 mm thick c-cut sapphire substrate. We measure spatially-resolved (1 μm resolution) reflectance $R(E)$ and transmittance $T(E)$ spectra of the crystals at normal incidence and at room temperature using a home-built setup (see Fig. S1 of the supplementary materials). Absorption spectra $A(E)$ are deduced from the $R(E)$ and $T(E)$ spectra as $A(E) = 1 - R(E) - T(E)$ [17,20]. Thicknesses of the crystals are initially estimated by comparing their observed colors under a microscope with colors calculated using the transfer matrix method (TMM) (see Fig. S13 and Fig. S14 of the supplementary materials) [15,22–24]. Crystal thicknesses are accurately measured using atomic force microscopy (AFM) (see Fig. S2 of the supplementary materials for examples).

In Fig. 1, we show the reflectance, transmittance and absorption spectra of monolayer (1L or 0.65 nm), 10 nm, and 110 nm thick $MoSe_2$, and 1L, 10 nm and 101 nm thick $WS_2$ as examples. In all cases, we notice well-known features corresponding to various excitonic resonances in the materials [27,28]. Additionally, after every resonance in the absorption spectrum, there is a step-like feature (Sommerfeld enhanced absorption step) [29], which is a classic signature of two dimensionality of these crystals, irrespective of their layer thicknesses. This thickness-independent two-dimensionality has been widely discussed in the literature on TMDCs, which has, for instance, led to the discovery of interlayer excitons in bilayers and multilayers [30]. In the present work, we do not focus on the exciton resonances, however, our target is to understand the integrated absorption (area under the absorption curve in the measured spectral region) as a function of layer thickness. As the flake thickness increases from 10 nm to about 100 nm, a notable enhancement in absorption is observed, accompanied by a reduction in transmittance. This is expected since thicker crystals interact much more strongly with the incident light.

Before going into any quantitative description, here we already notice an unusual behavior of reflectance of $MoSe_2$ where 110 nm thick crystal seems to reflect less strongly (Fig. 1(c)) compared to 10 nm case (Fig. 1(b)). This counterintuitive observation warrants detailed theoretical and experimental investigations of reflectance, transmittance and absorption of TMDCs over a broad region of thicknesses. Let us turn our attention to the absorption spectra first, which are relatively simpler to understand.

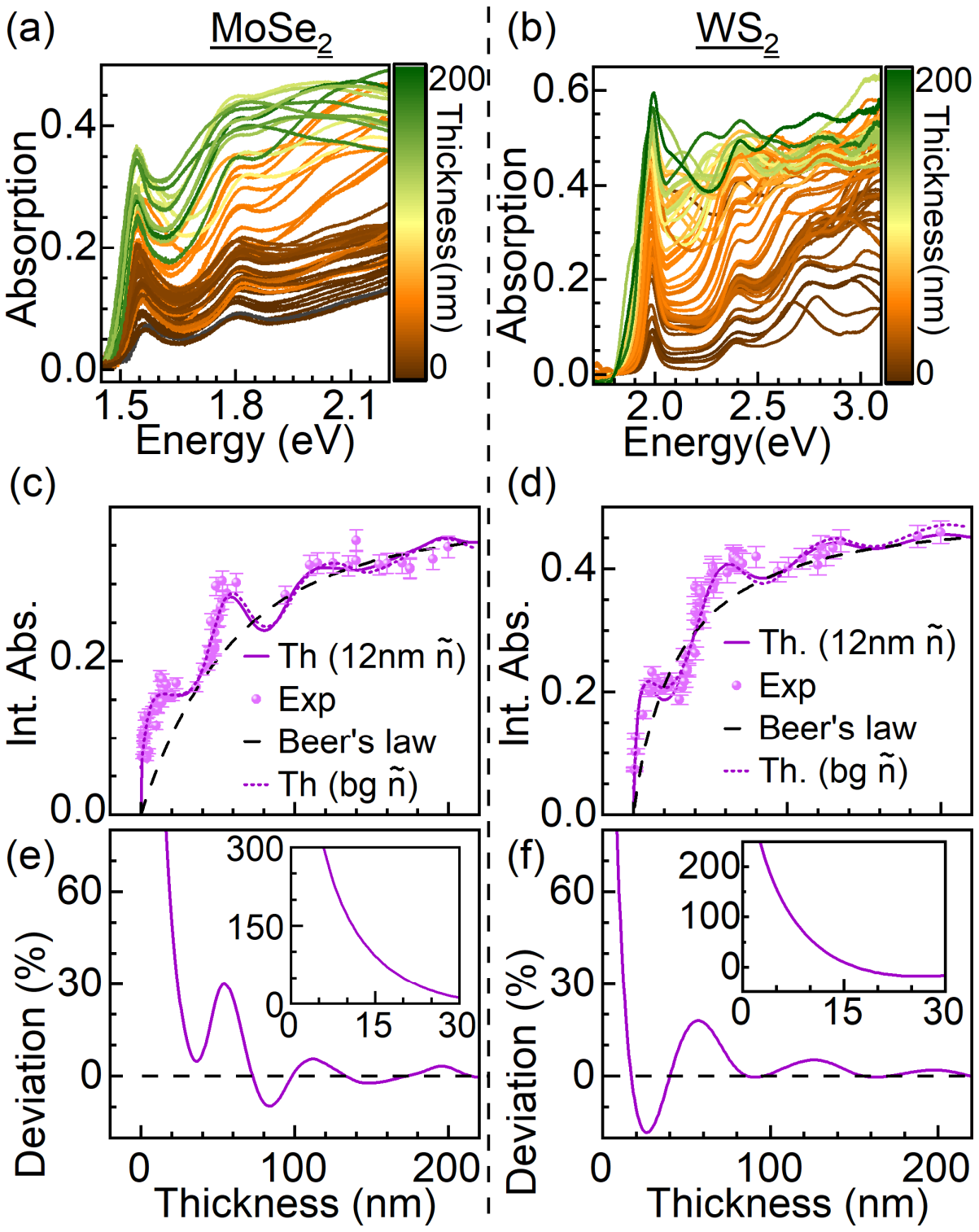

Figure 2. (a) and (b) are the experimental absorption spectra for $MoSe_2$ and $WS_2$ with varying thicknesses, ranging from monolayer (1L) to 200 nm for $MoSe_2$ and $WS_2$. Spheres in (c) and (d) are the experimental integrated absorption for the two materials with varying thickness. Dashed black lines are obtained using Beer's law. Solid lines are the theoretical curves obtained using complex refractive indices ($\tilde{n}$) of 12 nm thick crystals, while dotted lines are the curves obtained when exciton resonances are removed from $\tilde{n}$ (see Fig. S11 of supplementary materials). (e) and (f) represent the percentage deviation of the calculated integrated absorption using GTMM from the expected Beer's law i.e. $\left(100 \times \frac{A_{int}^{GTMM} - A_{int}^{Beer}}{A_{int}^{Beer}}\right)$. Insets depict the zoomed in deviation in absorption for up to 30 nm thickness.

Figures 2 (a) and (b) depict measured absorption spectra of $MoSe_2$ and $WS_2$ for thicknesses up to 200 nm, respectively. The spectral range for $MoSe_2$ is between $E_1 = 1.45$ eV to $E_2 = 2.2$ eV (bandwidth $\Delta E$ of 0.75 eV), while for $WS_2$, it is between $E_1 = 1.8$ eV to $E_2 = 3.1$ eV (bandwidth $\Delta E$ of 1.3 eV). The absorption approaches 0 below $E_1$. We realize that the experimentally measured absorption does not show a monotonic increase with thickness [Figs. 2(a) and 2(b)]. To verify this, we integrate the absorption spectra over experimental spectral regions as a function of layer thickness for both materials. We normalize the integrated absorption $A_{\mathrm{int}}$ with the spectral bandwidth i.e. the quantity $A_{\mathrm{int}}/\Delta E$ is plotted as solid spheres in Fig. 2(c) and 2(d) for $MoSe_2$ and $WS_2$ respectively, as a function of layer thickness. Notably, the monolayers and bilayers of both materials absorb ~7.5 % and ~10% of light, respectively. Strikingly, integrated absorption shows sudden jumps for a certain range of thicknesses e.g. $0 - 10$ nm (see Fig. S7 of the supplementary materials) and $30 - 55$ nm in both cases, while plateaus or even dips for other ranges (e.g. a plateau between $10 - 30$ nm and a dip between $55 - 85$ nm). The integrated absorption tends to saturate after 110 nm accompanied with subtle oscillatory behavior. Overall, this non-monotonic oscillatory behavior strongly deviates from the monotonically rising absorption predicted by the well-known Beer-Lambert law as detailed in the supporting information, and shown as dashed lines in Fig. 2(c) and 2(d). Drastic deviations occur at thicknesses below 10 nm in layered materials with high absorption coefficients, such as $WS_2$ (exceeding 50%) and $MoSe_2$ (exceeding 150%) as revealed in Figs. 2(e), 2(f). This is expected to significantly affect the absorption-based device design for future generation of ultrathin, flexible optoelectronics, as well as physics of light-matter interactions at the fundamental levels. The only experiments available in the literature find a monotonic, non-oscillatory behavior of integrated absorption in silicon covered with silica nanoparticles [13], $V_2O_5$ [14], $SiO_2$ [15], $MgF_2$ [15] or TiN [16] with thickness, contrary to our findings. This may be due to the authors measuring absorption at only a few thickness points or in a limited thickness range.

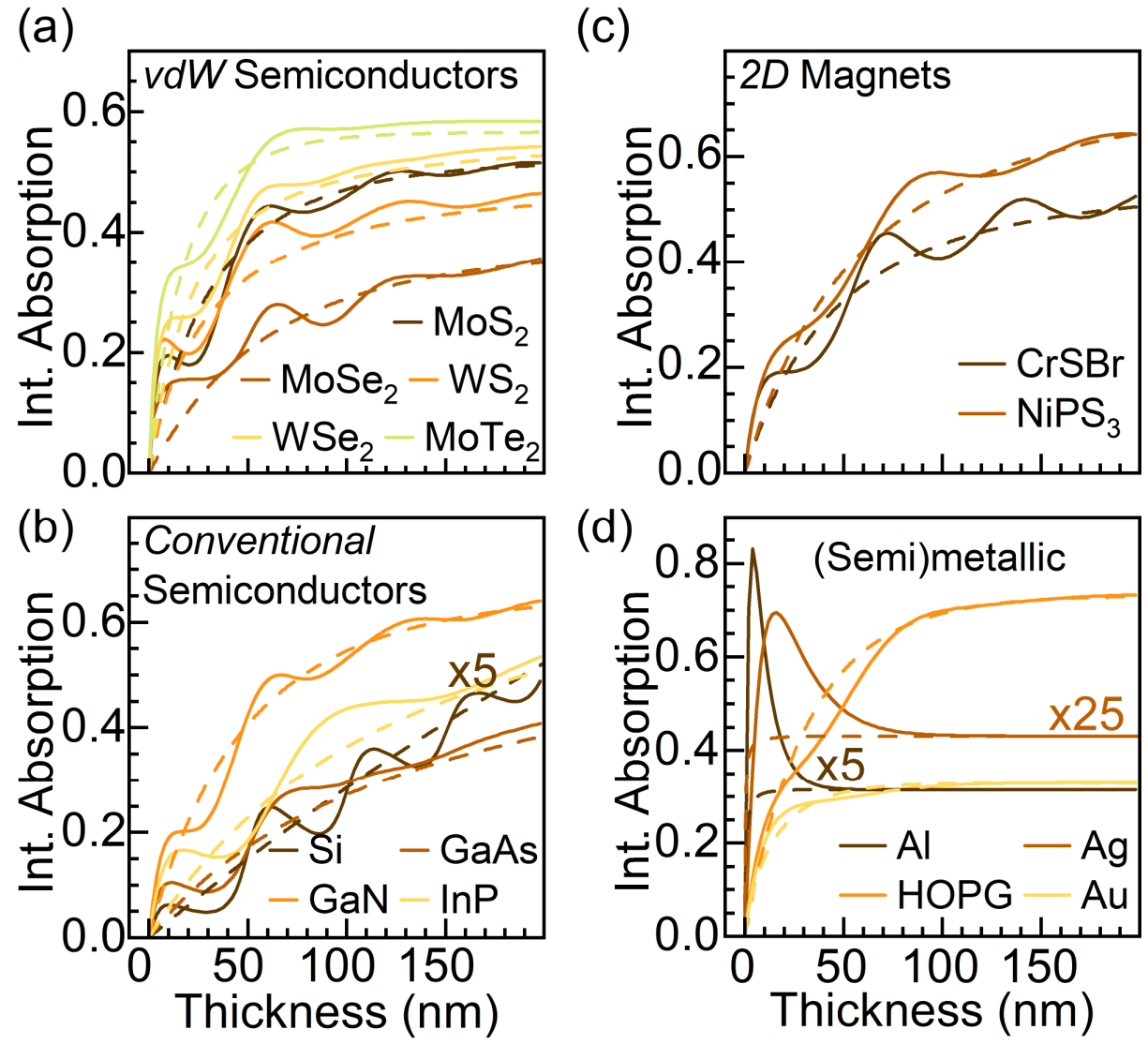

Figure 3. Solid and dashed lines show integrated absorption spectra using GTM method and Beer's law as discussed in supplementary materials (zoomed in view up to 20 nm available as Fig. S7). (a) vdW semiconductors: $MoS_2$, $MoSe_2$, $WS_2$, $WSe_2$ and $MoTe_2$ on 0.5mm thick sapphire;(b) freestanding conventional semiconductors i.e. Si (amplified by a factor of 5), GaAs, GaN and InP; (c) layered 2DMM: CrSBr and $NiPS_3$, on 0.5mm thick sapphire, (d) freestanding (semi)metallic materials i.e. Al (amplified by a factor of 5), Ag (amplified by a 25), Au and c-HOPG (Graphite on 0.5mm thick sapphire). The curves for $WS_2$ and $MoSe_2$ are from the present work, other curves are generated using refractive indices obtained from the literature mentioned in Table 1.

To understand the origin of the oscillatory features in the integrated absorption, we perform calculations based on the GTM method (see supplementary materials for details) [31,32]. Using this approach, we first calculate the photon-energy-dependent complex refractive indices $\tilde{n} = n + ik$ of $MoSe_2$ and $WS_2$ (see Fig. S11 of supplementary materials) as follows. We use experimentally measured reflectance $R(\lambda)$ and transmittance $T(\lambda)$ spectra of the 12 nm thick crystals of the two materials as inputs, while leaving the photon-energy-dependent dielectric functions $\tilde{\epsilon} = \epsilon_1 + i\epsilon_2 = (n + ik)^2$ as the free parameters. This thickness of 12 nm is chosen for our calculations because it can represent both regimes of ultrathin (1L limit) as well as the bulk reasonably well [33]. We note that the experimental results can be explained equally well using refractive indices of monolayers and bulk materials as well [34] (see Figs. S10 and S11 of the supplementary materials). Using these refractive indices as inputs in the GTM formalism, we compute absorption spectra $A(\lambda) = 1 - R(\lambda) - T(\lambda)$ as a function of thickness for $MoSe_2$ and $WS_2$ for the same spectral range as that of the experiment. Thereafter, we integrate area under these curves, normalize it by dividing with the spectral bandwidth $\Delta E$, and plot as solid lines in Figs. 2(c) and 2(d). The theoretical curves agree with the experimental data extremely well within the experimental error. To understand the effect of excitons on our calculations, we manually remove the excitonic features from $n$ and $k$ and obtain the background refractive indices as dashed lines shown in Fig. S10. The recalculated integrated absorption curves are depicted as dotted lines in Figs. 2(c) and 2(d) for $MoSe_2$ and $WS_2$, respectively. Strikingly, the overall qualitative behavior of the integrated absorption matches exceptionally well with the experimental data. Therefore, the oscillatory behavior of absorption must be universal to all solids, independent of the excitonic effects or any internal resonance of the system (see Figs. S10 and S12 of supplementary materials) or the choice of the substrate (see Figs. S9 and S17 of supplementary materials).

Table 1. Details of input parameters i.e. initial and final energies ($E_{int}$ and $E_{finl}$, respectively in eV) for computation, energy bandwidth ($\Delta E$ in eV) and reference used for refractive indices of materials using which integrated absorption has been calculated in Fig. 3.

| S. N. | Material | $E_{int}$ | $E_{finl}$ | $\Delta E$ | Reference |
|---|---|---|---|---|---|
| vdW Semiconductors/2D Magnets (Fig. 3a and 3c) | | | | | |
| 1 | 2H-$MoS_2$ | 1.8 | 2.7 | 0.9 | [39] |
| 2 | 2H-$MoSe_2$ | 1.45 | 2.2 | 0.75 | Present work |
| 3 | 2H-$WS_2$ | 1.8 | 3.1 | 1.3 | Present work |
| 4 | 2H-$WSe_2$ | 1.6 | 3 | 1.4 | [39] |
| 5 | 2H-$MoTe_2$ | 1.8 | 3.1 | 1.3 | [40] |
| 6 | CrSBr | 1.85 | 2.4 | 0.55 | [41] |
| 7 | $NiPS_3$ | 1.8 | 2.7 | 0.9 | [42] |
| Conventional semiconductors (Fig. 3b) | | | | | |
| 8 | Si | 1.4 | 2.7 | 1.3 | [43] |
| 9 | GaAs | 1.4 | 2.7 | 1.3 | [43] |
| 10 | GaN | 3 | 5 | 2 | [43] |
| 11 | InP | 1.4 | 2.7 | 1.3 | [44] |
| (Semi)metals (Fig. 3d) | | | | | |
| 12 | Au | 1.8 | 3.1 | 1.3 | [45] |
| 13 | Al | 1.8 | 3 | 1.2 | [46] |
| 14 | Ag | 1.8 | 3 | 1.2 | [45] |
| 15 | c-HOPG | 2 | 3.1 | 1.1 | [47] |

To further establish the universality of the effect, we calculate the integrated absorption of many layered materials, a few ‘conventional’ semiconductors namely Si, GaAs, GaN and InP, 2DMM such as CrSBr and $NiPS_3$, and (semi)metals i.e. Al, Ag, Graphite (c-HOPG) and Au as a function of layer thickness in Fig. 3, using parameters and the respective sources described in Table 1. The oscillatory behavior of the integrated absorption is obviously clear for all of the materials.

To gain a deeper understanding of the oscillatory integrated absorption, we take a closer look at the reflectance $R(\lambda)$, transmittance $T(\lambda)$ and absorption spectra $A(\lambda) = 1 - R(\lambda) - T(\lambda)$ as a function of layer thickness for $WS_2$ as an example. Figure 4 (a-c) presents the theoretical (solid lines, calculated using the GTM method) and experimental (circles) $R$, $T$ and $A$ for $WS_2$ at some specific wavelengths i.e. 400 nm to 700 nm in steps of 50 nm, as a function of layer thickness. In $R$ and $T$ curves, at a given wavelength, the expected oscillations due to interference taking place between the multiple interfaces of the sample i.e. air-crystal, crystal-substrate and substrate-air interfaces are visible [15,23,24]. The extrema appear for certain layer thicknesses where the conditions for interference are satisfied. For instance, for a freestanding layer (without substrate) of refractive index $n$, the maxima of reflectance (transmittance) would appear for $\mathrm{d} = (2\mathrm{m} - 1)\,\lambda/4\mathrm{n}$ ($\mathrm{d} = \mathrm{m}\lambda/2\mathrm{n}$) for an integer $m = 1,2,3 \ldots$ [20]. As expected from intuition, the extrema observed in $R$ and $T$ shift towards higher thicknesses for longer wavelengths. Our experimental results are in excellent agreement with the simulations, within the experimental errors of the size of the circles, as seen in Fig. 4. Fig. 4(c) reveals that the oscillatory behavior persists even in absorption vs layer thickness curves for all wavelengths. This is counterintuitive since it is usually assumed that the oscillatory effects in reflectance and transmittance cancel each other (energy conservation), being equal and opposite [35]. However, the amplitude of oscillations in absorption is not as pronounced as in reflectance or transmittance. Nonetheless, when integrated over all wavelengths in the spectral range, the oscillations persist in the integrated absorption in Fig. 4(d) which is seen experimentally. Interestingly, another form of ‘quantum’ universality in absorption spectral features of 2D layers has been reported earlier in 2D InAs membranes [36]. The authors showed that absorption spectra of InAs layers with thicknesses ranging from 3 – 19 nm display step-like features, with each step measured to be 1.6% of absorption. This phenomenon was theoretically explored across a wide range of quantum wells (group IV, III-V, II-VI and IV-VI) [37]. It was further extended to different dimensionalities including 0D quantum dots and 3D bulk semiconductors [38]. Complementarily, the present work does not deal with the spectral features. Instead, we explain universality of integrated absorption in solids over the classical electrodynamics regime irrespective of the spectral line shapes or presence of resonances.

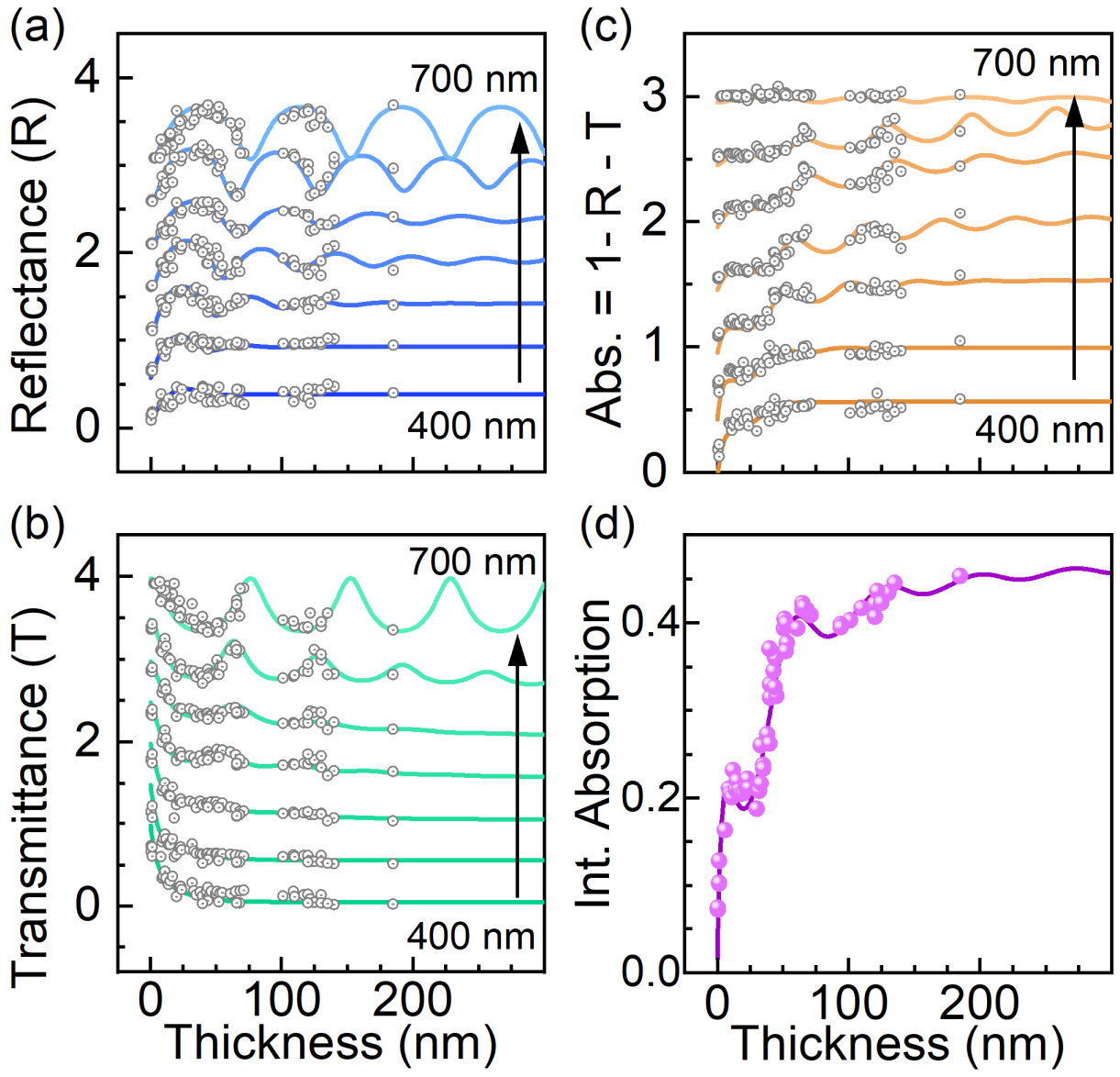

Figure 4. (a) to (c) represent calculated (solid lines) and experimental (circles) reflectance, transmittance and absorption of $WS_2$ on c-sapphire as a function of layer thickness, at selected wavelengths from 400 nm – 700 nm in steps of 50 nm. Plots are shifted in the y-axis for clarity, by 0.5 units successively. Fig 4(d) depicts the integrated absorption as a function of thickness for $WS_2$ where solid spheres are experimental data and solid line is theoretical curve calculated using GTM. The oscillatory behaviour from (c) persists in (d) as well.

In summary, the absorption integrated across the entire spectral range in solids exhibits oscillatory behavior with varying layer thickness, significantly deviating from the expected Beer-Lambert law. These findings are essential for deepening our understanding of light–matter interactions in solids and for enabling the design of optimized, absorption-driven ultrathin and flexible next-generation optoelectronic devices.

We acknowledge financial support from the following projects funded by the Government of India: NM-ICPS of the DST through the I-HUB Quantum Technology Foundation (Pune, India), DST Project No. CRG/2022/007008 of SERB, DST project No. ANRF/ARG/2025/002185/PS of ANRF, MoE-STARS project No. MoE-STARS/STARS-2/2023-0912, CEFIPRA CSRP Project No. 7104-2, VAIBHAV fellowship number INAE/DST-VF/2024/I/03, DST National Quantum Mission project No. DST/QTC/NQM/QMD/2024/4 (G), and IISER Pune IDeaS Scholarship. We thank T. S. Mahesh, Sandip Ghosh, G. V. Pavan Kumar, Shouvik Datta, Seema Sharma and Sourabh Dube for support with equipment while building the experimental setups, and Thorsten Deilmann for useful discussions.

[1] C. F. Klingshirn, *Semiconductor Optics* (Springer Berlin Heidelberg, Berlin, Heidelberg, 2012).

[2] S. Shree, I. Paradisanos, X. Marie, C. Robert, and B. Urbaszek, Guide to optical spectroscopy of layered semiconductors, Nat. Rev. Phys. **3**, 39 (2020).

[3] A. Arora, Magneto-optics of layered two-dimensional semiconductors and heterostructures : Progress and prospects, J. Appl. Phys. **129**, 120902 (2021).

[4] B. Huang, M. A. McGuire, A. F. May, D. Xiao, P. Jarillo-Herrero, and X. Xu, Emergent phenomena and proximity effects in two-dimensional magnets and heterostructures, Nat. Mater. **19**, 1276 (2020).

[5] J. H. Davies, *The Physics of Low-Dimensional Semiconductors* (Cambridge University Press, Cambridge, United Kingdom, 1998).

[6] J. Singh, *Electronic and Optoelectronic Properties of Semiconductor Structures* (Cambridge University Press, New York, 2003).

[7] T. Mueller and E. Malic, Exciton physics and device application of two-dimensional transition metal dichalcogenide semiconductors, Npj 2D Mater. Appl. **2**, 29 (2018).

[8] G. Wang, A. Chernikov, M. M. Glazov, T. F. Heinz, X. Marie, T. Amand, and B. Urbaszek, Colloquium : Excitons in atomically thin transition metal dichalcogenides, Rev. Mod. Phys. **90**, 021001 (2018).

[9] W. Liu et al., Flexible solar cells based on foldable silicon wafers with blunted edges, Nature **617**, 717 (2023).

[10] Z. Hu, S. Wang, J. Lynch, and D. Jariwala, Tandem Photovoltaics from 2D Transition Metal Dichalcogenides on Silicon, ACS Photonics **11**, 4616 (2024).

[11] F. Bozheyev, Advancement of transition metal dichalcogenides for solar cells: a perspective, J. Mater. Chem. A **11**, 19845 (2023).

[12] M. Bernardi, M. Palummo, and J. C. Grossman, Extraordinary Sunlight Absorption and One Nanometer Thick Photovoltaics Using Two-Dimensional Monolayer Materials, Nano Lett. **13**, 3664 (2013).

[13] P. Banerjee, A. B. Roy, A. Nandi, S. Das, A. Kundu, S. Mukherjee, H. Saha, and S. M. Hossain, Light-trapping scheme using silica spheres on ultrathin c-silicon absorber: transition from antireflection coating to whispering gallery resonator, Appl. Phys. A **128**, 481 (2022).

[14] S. A. Aly, S. A. Mahmoud, N. Z. El-Sayed, and M. A. Kaid, Study on some optical properties of thermally evaporated $V_2O_5$ films, Vacuum **55**, 159 (1999).

[15] O. Stenzel and M. Ohlídal, editors , *Optical Characterization of Thin Solid Films*, Vol. 64 (Springer International Publishing, Cham, 2018).

[16] T. S. Luk, G. Xu, W. Ross, J. N. Nogan, E. A. Scott, S. Ivanov, O. Niculescu, O. Mitrofanov, and C. T. Harris, Maximal absorption in ultrathin TiN films for microbolometer applications, Appl. Phys. Lett. **121**, 234101 (2022).

[17] M. Fox, *Optical Properties of Solids* (OUP Oxford, 2010).

[18] J. I. Pankove, Optical process in semiconductors, Dover Publ. Inc. **119**, 450 (1975).

[19] T. G. Mayerhöfer, S. Pahlow, and J. Popp, The Bouguer-Beer-Lambert Law: Shining Light on the Obscure, ChemPhysChem **21**, 2029 (2020).

[20] E. Hecht, *Optics*, 4th ed. (Pearson Addison Wesley, Reading, 2001).

[21] M. Born, E. Wolf, A. B. Bhatia, P. C. Clemmow, D. Gabor, A. R. Stokes, A. M. Taylor, P. A. Wayman, and W. L. Wilcock, *Principles of Optics* (Cambridge University Press, 1999).

[22] P. Blake, E. W. Hill, A. H. Castro Neto, K. S. Novoselov, D. Jiang, R. Yang, T. J. Booth, and A. K. Geim, Making graphene visible, Appl. Phys. Lett. **91**, (2007).

[23] H. Anders, *Thin Films in Optics* (Focal Press, London, 1967).

[24] S. A. Furman and A. V. Tikhonravov, *Basics of Optics of Multilayer Systems* (Edition Frontieres, Gif-Sur-Yvette, 1992).

[25] C. E. Stevens, T. Stroucken, A. V. Stier, J. Paul, H. Zhang, P. Dey, S. A. Crooker, S. W. Koch, and D. Karaiskaj, Superradiant coupling effects in transition-metal dichalcogenides, Optica **5**, 749 (2018).

[26] A. Castellanos-Gomez, M. Buscema, R. Molenaar, V. Singh, L. Janssen, H. S. J. van der Zant, and G. a Steele, Deterministic transfer of two-dimensional materials by all-dry viscoelastic stamping, 2D Mater. **1**, 011002 (2014).

[27] A. Arora, K. Nogajewski, M. Molas, M. Koperski, and M. Potemski, Exciton band structure in layered $MoSe_2$: from a monolayer to the bulk limit, Nanoscale **7**, 20769 (2015).

[28] M. R. Molas, K. Nogajewski, A. O. Slobodeniuk, J. Binder, M. Bartos, and M. Potemski, Optical response of monolayer, few-layer and bulk tungsten disulfide, Nanoscale **9**, 13128 (2017).

[29] S. L. Chuang, *Physics of Optoelectronic Devices* (John Wiley & Sons, Inc., New York, 1995).

[30] A. Arora et al., Interlayer excitons in a bulk van der Waals semiconductor, Nat. Commun. **8**, 639 (2017).

[31] E. Centurioni, Generalized matrix method for calculation of internal light energy flux in mixed coherent and incoherent

multilayers, Appl. Opt. **44**, 7532 (2005).

[32] R. Santbergen, A. H. M. Smets, and M. Zeman, Optical model for multilayer structures with coherent, partly coherent and incoherent layers, Opt. Express **21**, A262 (2013).

[33] Y. Niu et al., Thickness-Dependent Differential Reflectance Spectra of Monolayer and Few-Layer $MoS_2$, $MoSe_2$, $WS_2$ and $WSe_2$, Nanomaterials **8**, 725 (2018).

[34] Y. Li, A. Chernikov, X. Zhang, A. Rigosi, H. M. Hill, A. M. van der Zande, D. A. Chenet, E.-M. Shih, J. Hone, and T. F. Heinz, Measurement of the optical dielectric function of monolayer transition-metal dichalcogenides: $MoS_2$, $MoSe_2$, $WS_2$, and $WSe_2$, Phys. Rev. B **90**, 205422 (2014).

[35] H. A. Macleod, *Thin-Film Optical Filters, Fifth Edition* (CRC Press, 2017).

[36] H. Fang, H. A. Bechtel, E. Plis, M. C. Martin, S. Krishna, E. Yablonovitch, and A. Javey, Quantum of optical absorption in two-dimensional semiconductors, Proc. Natl. Acad. Sci. **110**, 11688 (2013).

[37] M. Lannoo, P. T. Prins, Z. Hens, D. Vanmaekelbergh, and C. Delerue, Universality of optical absorptance quantization in two-dimensional group-IV, III-V, II-VI, and IV-VI semiconductors, Phys. Rev. B **105**, 035421 (2022).

[38] P. T. Prins, M. Alimoradi Jazi, N. A. Killilea, W. H. Evers, P. Geiregat, W. Heiss, A. J. Houtepen, C. Delerue, Z. Hens, and D. Vanmaekelbergh, The Fine-Structure Constant as a Ruler for the Band-Edge Light Absorption Strength of Bulk and Quantum-Confined Semiconductors, Nano Lett. **21**, 9426 (2021).

[39] C. Hsu, R. Frisenda, R. Schmidt, A. Arora, S. M. de Vasconcellos, R. Bratschitsch, H. S. J. van der Zant, and A. Castellanos-Gomez, Thickness-Dependent Refractive Index of 1L, 2L, and 3L $MoS_2$ , $MoSe_2$ , $WS_2$ , and $WSe_2$, Adv. Opt. Mater. **7**, 1900239 (2019).

[40] B. Munkhbat, P. Wróbel, T. J. Antosiewicz, and T. O. Shegai, Optical Constants of Several Multilayer Transition Metal Dichalcogenides Measured by Spectroscopic Ellipsometry in the 300–1700 nm Range: High Index, Anisotropy, and Hyperbolicity, ACS Photonics **9**, 2398 (2022).

[41] G. Ermolaev et al., Giant optical anisotropy in CrSBr from giant exciton oscillator strength, ArXiv Preprint at arXiv:2509.18866 (2025).

[42] P. G. Zotev et al., Van der Waals Materials for Applications in Nanophotonics, Laser Photon. Rev. **17**, (2023).

[43] D. E. Aspnes and A. A. Studna, Dielectric functions and optical parameters of Si, Ge, GaP, GaAs, GaSb, InP, InAs, and InSb from 1.5 to 6.0 eV, Phys. Rev. B **27**, 985 (1983).

[44] S. Adachi, Optical dispersion relations for GaP, GaAs, GaSb, InP, InAs, InSb, $Al_xGa_{1-x}As$, and $In_{1x}Ga_xAs_yP_{1-y}$, J. Appl. Phys. **66**, 6030 (1989).

[45] S. Babar and J. H. Weaver, Optical constants of Cu, Ag, and Au revisited, Appl. Opt. **54**, 477 (2015).

[46] A. D. Rakić, Algorithm for the determination of intrinsic optical constants of metal films: application to aluminum, - Appl. Opt. **34**, 4755 (1995).

[47] B. Song, H. Gu, S. Zhu, H. Jiang, X. Chen, C. Zhang, and S. Liu, Broadband optical properties of graphene and HOPG investigated by spectroscopic Mueller matrix ellipsometry, Appl. Surf. Sci. **439**, 1079 (2018).

Supplemental Material

# Universal Thickness-Dependent Absorption in Solids at the Nanoscale

*Bhumika Chauhan[1], Nikhil Singh[1], Subhrajit Dalai[1], Abhisek Saidarsan[1], Sayantan Patra[1], Sourabh Jain[1], Aparna Deshpande[1], Ashish Arora[1,*]*

*[1]Department of Physics, Indian Institute of Science Education and Research, Dr. Homi Bhabha Road, 411008 Pune, India*

Email: ashish.arora@iiserpune.ac.in

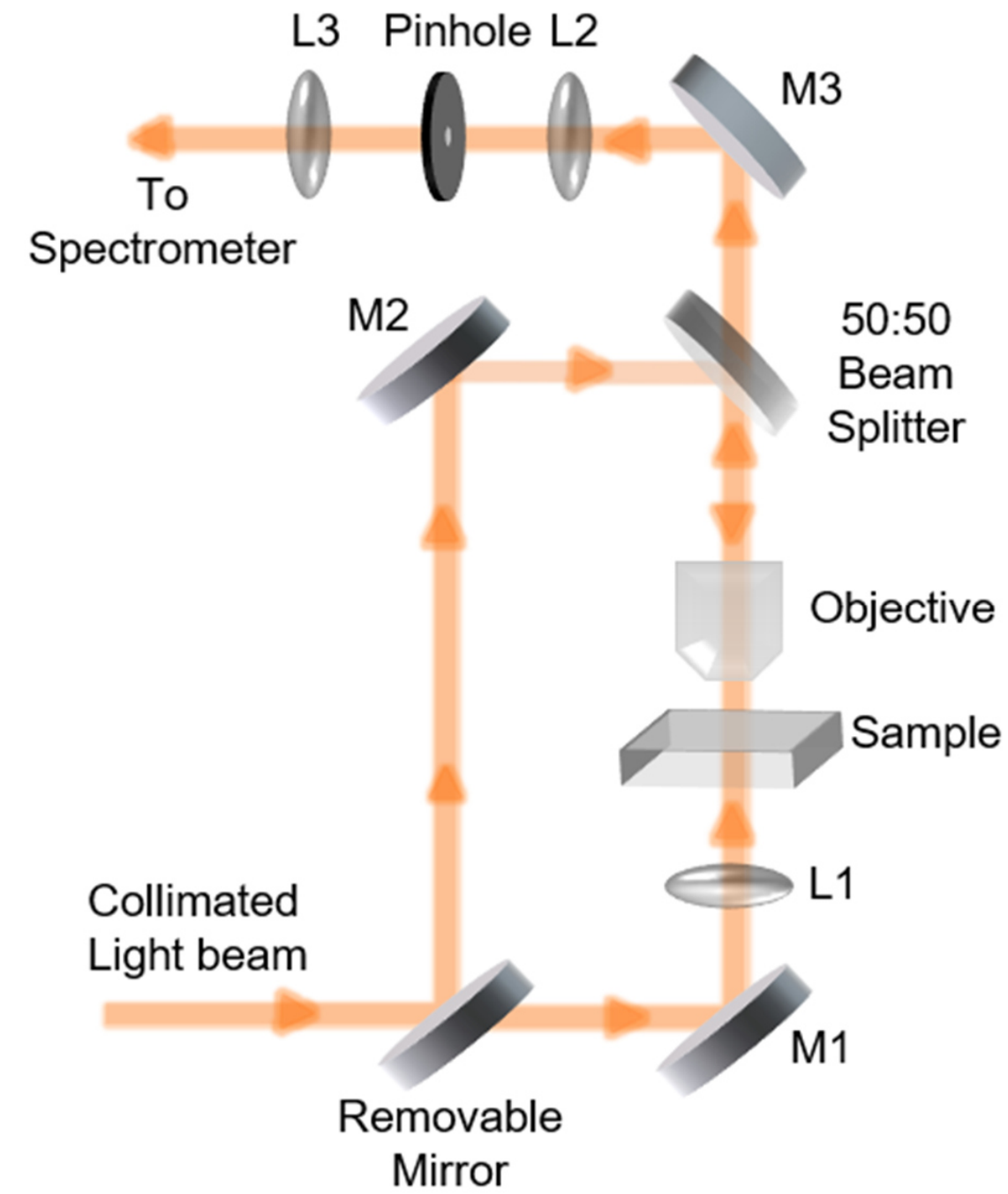

**Figure S1. Experimental Setup.**

1) *Reflection mode:* Broadband collimated light beam from a Xe arc lamp or tungsten Halogen lamp is reflected from the removable mirror, undergoes reflection from the M2 and 50:50 beam splitter and is focused on the sample using 50x objective lens. The reflected light from the sample passes through the 50:50 beam splitter, reflects from M3, and focused on pinhole of 20μm diameter using L2, while L3 collimates it again. The pinhole selects signal from ~1μm region of the sample.

2) *Transmission mode:* Collimated light from the light source is reflected from M1 and is focused on the sample by L1. Afterwards, it is the same as reflection mode. Light after L3 is wavelength dispersed after passing through a 328 mm focal length spectrometer, and is detected using a Peltier-cooled CCD.

### Absorption from Beer-Lambert law

### Free Standing Layer: *Air-crystal -Air* system

Consider a solid slab of thickness $d$, with an absorption coefficient $\alpha$, bounded by air on both sides. This configuration constitutes an air-solid-air system as illustrated in Figure S3.

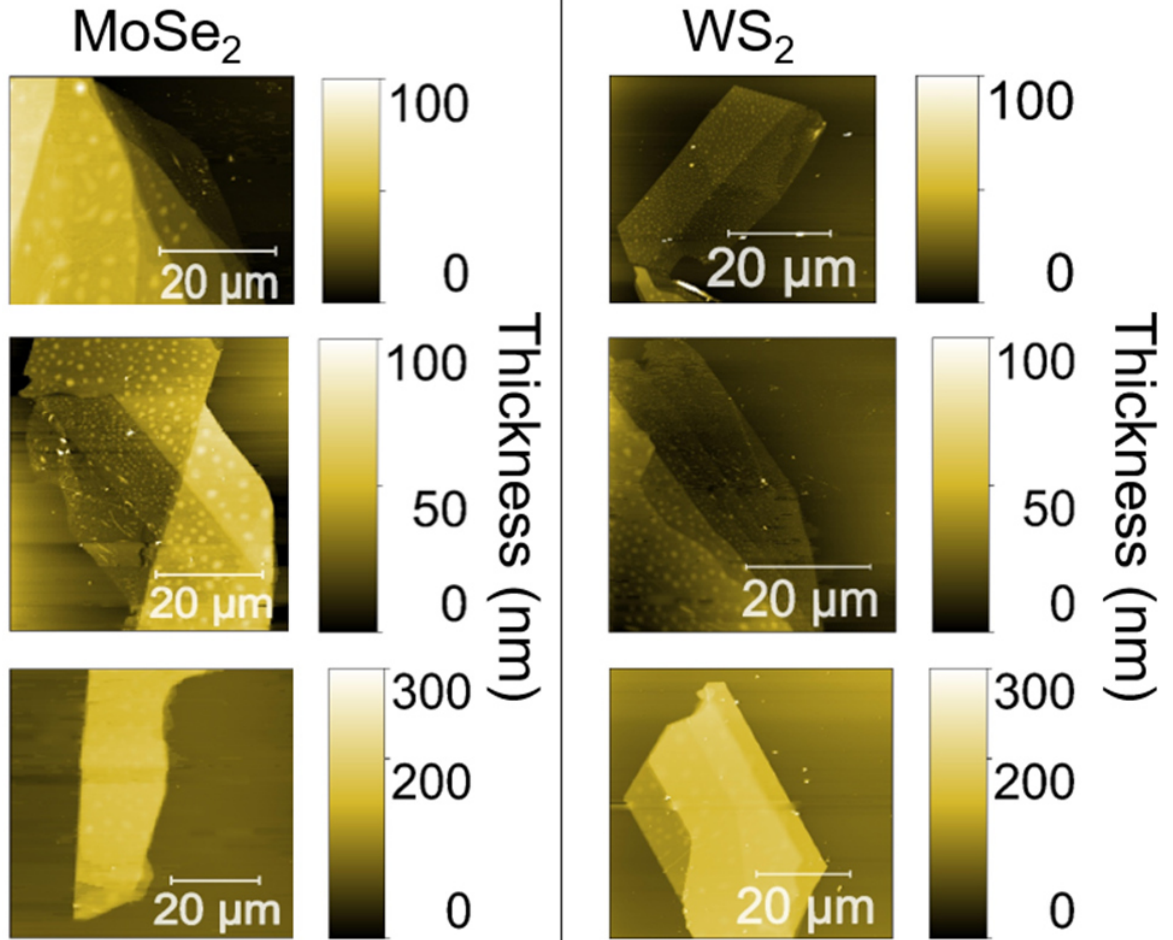

**Figure S2.** Tapping-mode atomic force microscopy (AFM) height maps of representative mechanically exfoliated $MoSe_2$ (left) and $WS_2$ (right) crystals on 0.5mm thick c-sapphire substrate. These images correspond to the same regions shown in the optical micrographs presented in Fig. 1. Similar measurements were performed extensively across multiple flakes to ensure reproducibility and consistency of the observed thickness-dependent behavior of absorption reported in this work.

At the *air-solid* interface, consider an infinitesimally thin region of thickness $dx$. Since absorption within this region is negligible, energy conservation suggests

$$R_o + T_o = 1 \tag{1}$$

where, $R_o$ and $T_o$ are the reflectance and transmittance at the interfaces $x = 0$ and $x = d$, respectively.

Generally, within a solid, the transmitted intensity $I(x)$ at location $x$ can be expressed in terms of the intensity entering the solid $I_o$ at $x = 0$, using Beer–Lambert law as [1]

$$I(x) = I_o e^{-\alpha x} \tag{2}$$

A reflection/transmission (R/T) at every successive interface (including the first interface) leads to a new $I_0$ in Eq. 2, which takes the losses due to R and T at the interface into account. Therefore, the total reflectance of the system, $R$, can be obtained by considering all internally reflected rays that eventually exit backwards through the $x = 0$ *air-solid* interface as follows [1–3]

$$R = R_o + (1 - R_o)^2 R_o e^{-2\alpha d} + (1 - R_o)^2 R_0^3 e^{-4\alpha d} + \cdots$$

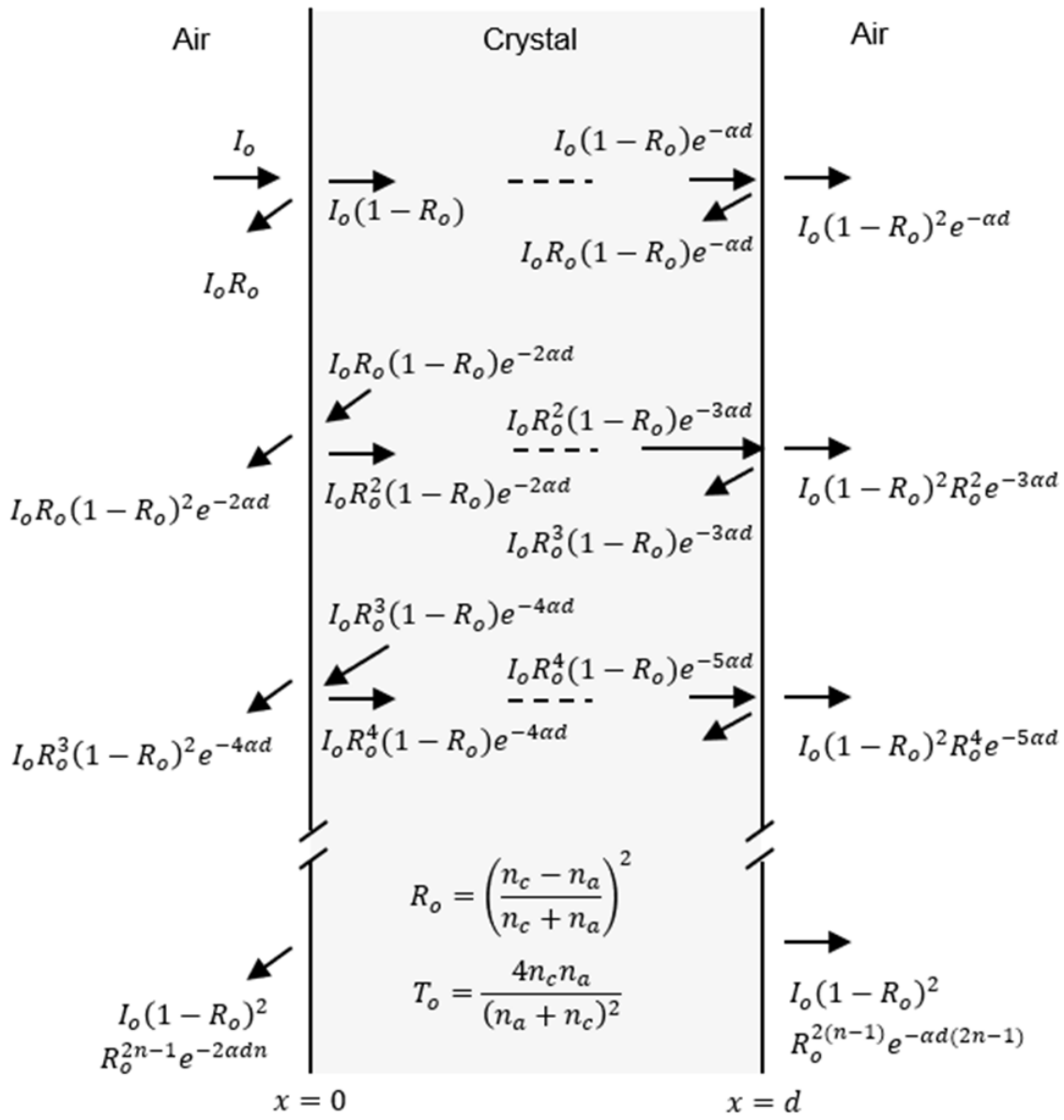

**Figure S3.** The air-crystal-air (freestanding) system where energy losses of the incident beam due to reflection losses at the interfaces, and attenuation due to Beer-Lambert law in the medium, has been explicitly accounted for. Here, $I_o$ is the intensity of the incident beam, $\alpha$ is the absorption coefficient of the crystal, and $R_o$ is the reflectance at the *air-crystal* interface, calculated using $n_a$ and $n_c$, which are the refractive indices of air and the crystal respectively.

$$= R_o + (1 - R_o)^2 e^{-2\alpha d} R_o \sum_{n=0}^{\infty} R_o^{2n} e^{-2\alpha dn}$$

$$= R_o + \frac{(1 - R_o)^2 R_o}{1 - R_o^2 e^{-2\alpha d}} e^{-2\alpha d} \tag{3}$$

Similarly, by considering all rays that eventually exit through $x = d$ *solid-air* interface, the overall transmittance of the system, T, can be written by summing over these contributions as follows [1–3]

$$T = (1 - R_o)^2 e^{-\alpha d} + (1 - R_o)^2 R_o^2 e^{-3\alpha d} + (1 - R_o)^2 R_o^4 e^{-5\alpha d} \ldots$$

$$= (1 - R_o)^2 e^{-\alpha d} \sum_{n=0}^{\infty} e^{-2\alpha dn} R_o^{2n}$$

$$= \frac{(1 - R_o)^2}{1 - R_o^2 e^{-2\alpha d}} e^{-\alpha d} \tag{4}$$

The effective absorption of the system, $A$, can therefore be computed using Eq. (3) and Eq. (4) as follows [1–3]

$$A = 1 - R - T$$

$$= 1 - \left( R_o + \frac{(1 - R_o)^2 R_o}{1 - R_o^2 e^{-2\alpha d}} e^{-2\alpha d} \right) - \left( \frac{(1 - R_o)^2}{1 - R_o^2 e^{-2\alpha d}} e^{-\alpha d} \right)$$

$$= \frac{(1 - R_o)(1 - e^{-\alpha d})}{(1 - R_0 e^{-\alpha d})} \tag{5}$$

***Air-Crystal-Substrate-Air* system:** Extending the approach described in the preceding section, the effective reflectance and transmittance (and thereby absorption) of an *air–crystal–substrate–air* system can be obtained by accounting for multiple internal reflections at each interface, and attenuation with each medium, as shown in Fig. S4.

Reflectance coefficients at the *air–crystal*, *crystal–substrate*, and *substrate–air* interfaces in Fig. S4 are denoted as $R_o^{ac}$, $R_o^{cs}$ and, $R_o^{sa}$, respectively. Absorption coefficients of the crystal and substrate are represented by $\alpha_c$, and $\alpha_s$, and their respective thicknesses by $d_c$ and $d_s$.

We define a parameter, $\gamma$,

$$\gamma = R_{sa} + \frac{(1 - R_{sa})^2 R_{cs} \mathrm{e}^{-2\alpha_s d_s}}{1 - R_{sa} R_{cs}\, e^{-2\alpha_s d_s}} \tag{6}$$

in terms of which, the total reflectance $R_{total}$, and total transmittance $T_{total}$, of the system can be written as follows:

$$R_{total} = R_{ac} + \frac{(1 - R_{ac})^2 \gamma e^{-2\alpha_c d_c}}{1 - R_{ac} \gamma e^{-2\alpha_c d_c}} \tag{7}$$

$$T_{total} = \frac{(1 - R_{ac})(1 - R_{cs})(1 - R_{sa}) e^{-(\alpha_c d_c + \alpha_s d_s)}}{(1 - R_{ac} \gamma e^{-2\alpha_c d_c})(1 - R_{sa} R_{cs} e^{-2\alpha_s d_s})} \tag{8}$$

Therefore, the total absorption of the system can be computed using equations (7) and (8) as follows:

$$A_{total} = 1 - R_{total} - T_{total} \tag{9}$$

This expression was used to calculate the absorption from Beer-Lambert law for our system, using complex refractive indices (with excitonic features) and plotted in Fig.2 (c) and (d) in the main manuscript.

**Generalized Transfer matrix model:**

In many cases, the substrate thicknesses are much larger than the coherence length of the light source (e.g. tungsten halogen or xenon arc lamps) used for spectroscopy. To account for the effect of substrates properly, the *conventional* transfer-matrix method such as the one discussed in Ref. [4] fails. It is because, there is no provision for introducing incoherence in this method. A complete transfer-matrix calculation including a substrate results in strong oscillations in absorption spectra such as in Fig. S6 (a and b) for $MoSe_2$ and $WS_2$ kept on 0.5 mm thick c-sapphire substrate. Therefore, in the present work, we have performed our calculations using an improved version of the transfer-matrix method, called generalized transfer-matrix method [5,6]. In this method, the substrates can be set as incoherent and the results are much closer to reality (Fig. S6 (c and d)). Here we summarize the method as follows.

Let us consider a multilayer stack of $i$ films which makes $i$ +1 interfaces as shown in Fig. S5. The semi-infinite media on the leftmost ($0^{th}$) and on the rightmost ($i + 1^{th}$) sides are considered as air or vacuum. Linearly polarized light passing through this structure can be described with the transfer-matrix formalism:

$$\begin{bmatrix} E_0^+ \\ E_0^- \end{bmatrix} = M \begin{bmatrix} E_{i+1}^+ \\ E_{i+1}^- \end{bmatrix} \tag{10}$$

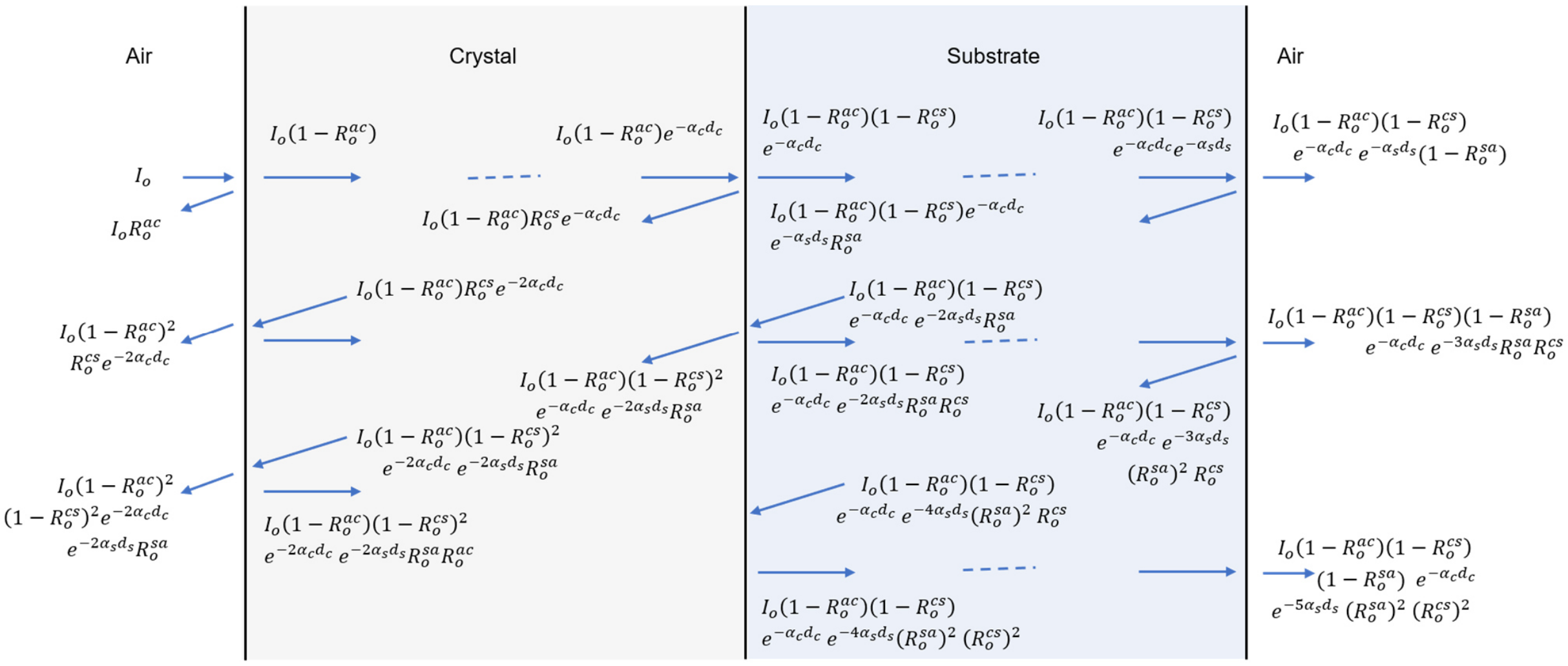

**Figure S4:** The *air-crystal-substrate-air* system where energy losses of the incident beam due to reflection losses at each interface, and attenuation due to Beer-Lambert law in the respective media, has been explicitly accounted for. Here, $I_o$ is the intensity of the incident beam. Reflection coefficients at the *air–crystal*, *crystal–substrate*, and *substrate–air* interfaces are denoted as $R_o^{ac}$, $R_o^{cs}$ and, $R_o^{sa}$, respectively. Absorption coefficients of the crystal and substrate are represented by $\alpha_c$, and $\alpha_s$, and their respective thicknesses by $d_c$ and $d_s$. Note that only the leading order terms have been shown in the illustration.

where $E_0^+, E_{i+1}^+$ and $E_0^-, E_{i+1}^-$ are the forward and backward propagating electric field amplitudes at $0^{th}$ and $i+1^{th}$ interfaces. $M$ is a $2 \times 2$ transfer matrix $M = \begin{pmatrix} M_{11} & M_{12} \\ M_{21} & M_{22} \end{pmatrix}$ given by

$$M = I_{01}.L_1.I_{12} \ldots L_{\mathrm{i}}.I_{\mathrm{i(i+1)}} \tag{11}$$

where $I_{\mathrm{ij}}$ are $2 \times 2$ 'interface' matrices which connect the fields at the $i^{th}$ and $j^{th}$ boundaries, while $L_i$ are the propagation matrices which describe the change in phase while wave propagates through the $i^{th}$ layer. Both of these matrices are described in detail as follows.

$$\boldsymbol{I}_{\mathrm{ij}} = \frac{1}{t_{\mathrm{ij}}} \begin{bmatrix} 1 & r_{\mathrm{ij}} \\ r_{\mathrm{ij}} & 1 \end{bmatrix} \tag{12}$$

where $t_{ij}$ and $r_{ij}$ are the complex Fresnel's transmission and reflection coefficients at the interface $ij$ for normal incidence (consider $s$ polarization):

$$r_{\mathrm{ij}} = \frac{n_{\mathrm{i}} - n_{\mathrm{j}}}{n_{\mathrm{j}} + n_{\mathrm{i}}} \text{ and } t_{\mathrm{ij}} = \frac{2\, n_{\mathrm{i}}}{n_{\mathrm{i}} + n_{\mathrm{j}}} \tag{13}$$

with $n_i$ and $n_j$ as the complex refractive indices of $i^{th}$ and $j^{th}$ films.

The matrix L (propagation matrix) is defined as

$$\boldsymbol{L}_{\mathrm{i}} = \begin{bmatrix} e^{-i\beta_{\mathrm{i}}} & 0 \\ 0 & e^{i\beta_{\mathrm{i}}} \end{bmatrix} \tag{14}$$

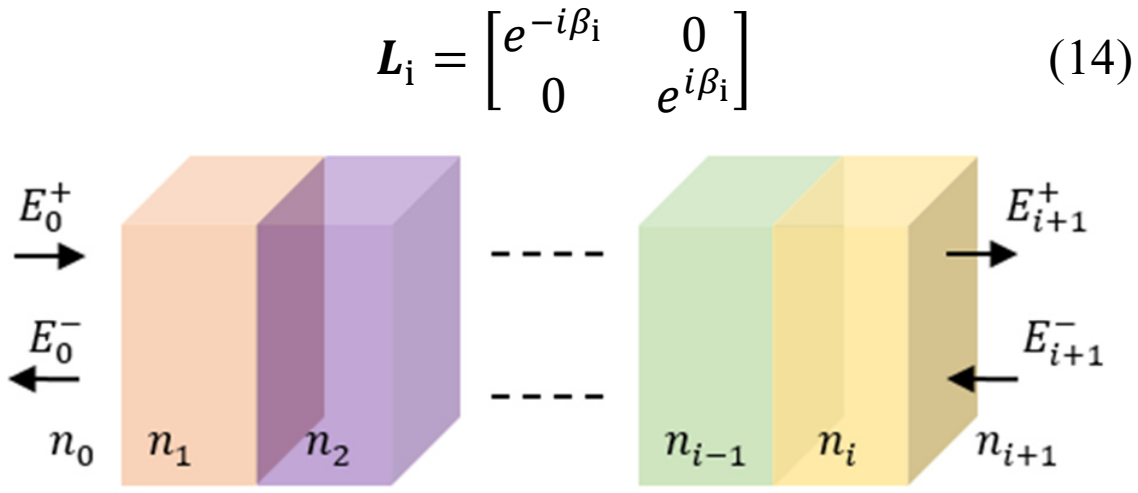

**Figure S5.** Schematic of the multilayer optical model. A dielectric stack is represented with refractive index $n_i$ and thickness $d_i$.This model is used to describe how light propagates through thin films and to calculate reflectance, transmittance, and absorption.

where $\beta_i$ is the phase shift for the wave passing through the film $i$ and is defined as

$$\beta_{\mathrm{i}} = \frac{2\pi n_{\mathrm{i}} d_{\mathrm{i}}}{\lambda} \tag{15}$$

where λ is the wavelength of the incident light, $n_i$ and $d_i$ are the complex refractive index and the thickness of $i^{th}$ film, respectively.

Now we proceed to calculate the final reflectance and transmittance coefficients through the multilayered system. Let us find the reflectance and transmittance amplitudes first. From (10),

$$\begin{bmatrix} E_0^+ \\ E_0^- \end{bmatrix} = \begin{bmatrix} M_{11} & M_{12} \\ M_{21} & M_{22} \end{bmatrix} \begin{bmatrix} E_{i+1}^+ \\ 0 \end{bmatrix} = \begin{bmatrix} M_{11} E_{i+1}^+ \\ M_{21} E_{i+1}^+ \end{bmatrix} \tag{16}$$

with $E_{i+1}^- = 0$ (since there is no light incident back from the right side), reflection and transmission amplitudes are

$$r = \frac{E_0^-}{E_0^+} = \frac{M_{21}}{M_{11}} \text{ and } t = \frac{E_{i+1}^+}{E_0^+} = \frac{1}{M_{11}} \tag{17}$$

The reflectance and transmittance are, therefore, given as:

$$R = r^* r = \left| \frac{M_{21}}{M_{11}} \right|^2 \tag{18}$$

$$T = \left( \frac{n_{i+1}}{n_0} \right) t^* t = \left( \frac{n_{i+1}}{n_0} \right) \left| \frac{1}{M_{11}} \right|^2 \tag{19}$$

where $r^*$ and $t^*$ are the complex conjugates of the reflection and transmission amplitudes.

So far, light passing through every layer in the multilayered structure behaves coherently i.e. light accumulates phase as it passes through the different layers. However, it is problematic since the substrates, which are usually very thick compared to the coherence length of the light source, cannot be treated in this fashion. This is the reason we notice strong oscillations in Fig. S6(a) and (b) riding over the calculated spectra, which are absent in the experiment. For such thick layers, the relative phase between forward and backward-propagating waves inside that layer becomes effectively random. To introduce this *decoherence* effect for the substrate, we calculate $r$ and $t$ for several $\beta_{\mathrm{i}} = \frac{2\pi n_{\mathrm{i}} d_{\mathrm{i}}}{\lambda} + \phi_m$ and average those over [6]. Here $\phi_m's$ are equidistant phases between 0 and $2\pi$, separated by $2\pi/m$, $m$ being an integer. For instance, we use $m = 8$, where the $\phi_m's$ are $0, \frac{\pi}{4}, \frac{\pi}{2}, \ldots, 2\pi$. Our calculation reveals that the curves obtained through this method are extremely close to the reality [see Figs. S6(c) and (d)]. Therefore, we use this method everywhere in the present work.

It is important to note that the GTMM approach explicitly accounts for the interference effects between the light waves in each medium, and, thereby, the formation of standing waves within the system, whereas the Beer-Lambert-based ray tracing method does not.

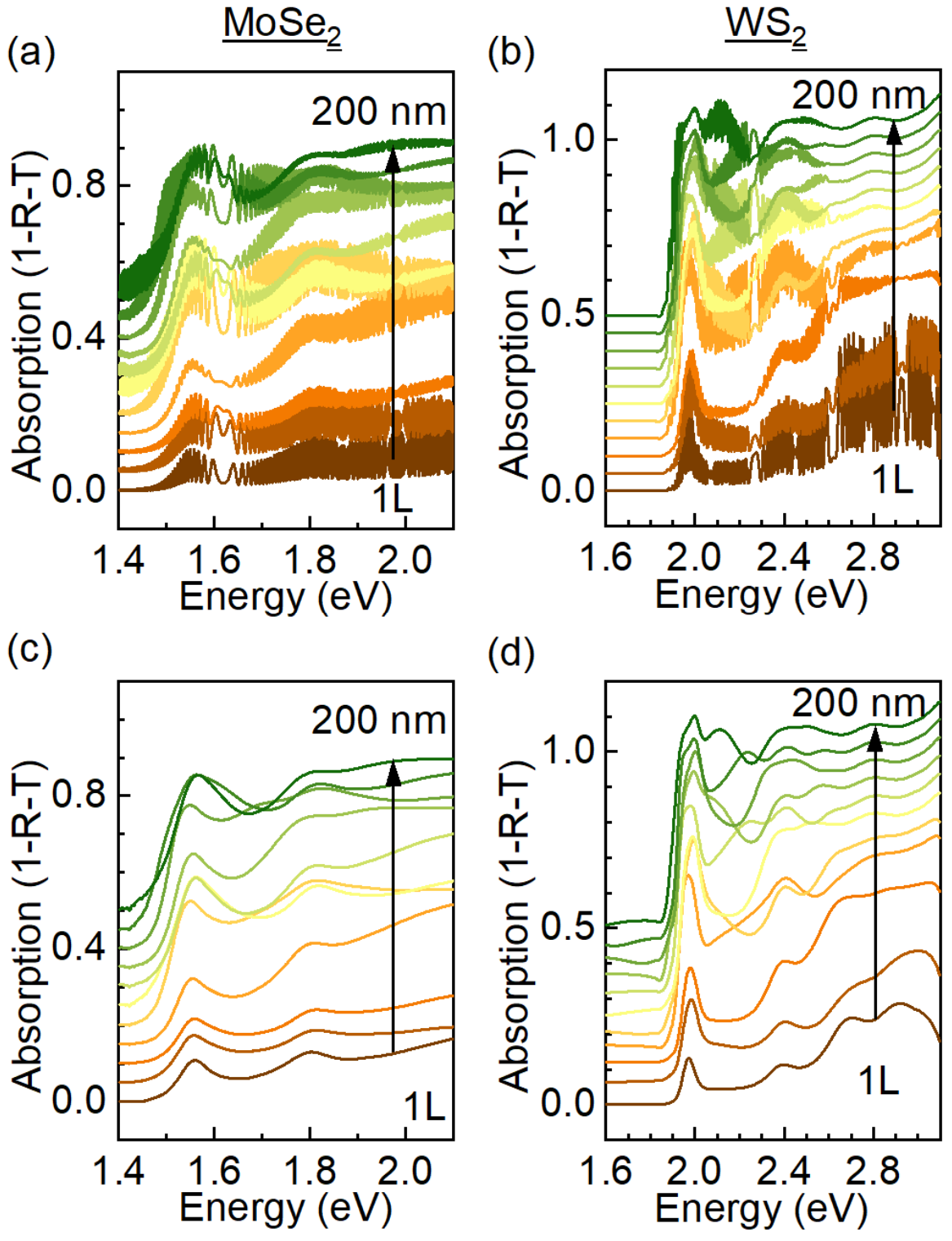

**Figure S6.** Calculated absorption spectra using *conventional* transfer-matrix method in (a) and (b) for $MoSe_2$ and $WSe_2$, respectively, and using *generalized* transfer-matrix method in (c) and (d) where the TMDC crystals are kept on 0.5 mm thick c-sapphire substrate. The spectra are calculated for TMDC thicknesses ranging from 1L to 200 nm in steps of 20 nm. The successive spectra are shifted along the y-axes by 0.05 for clarity.

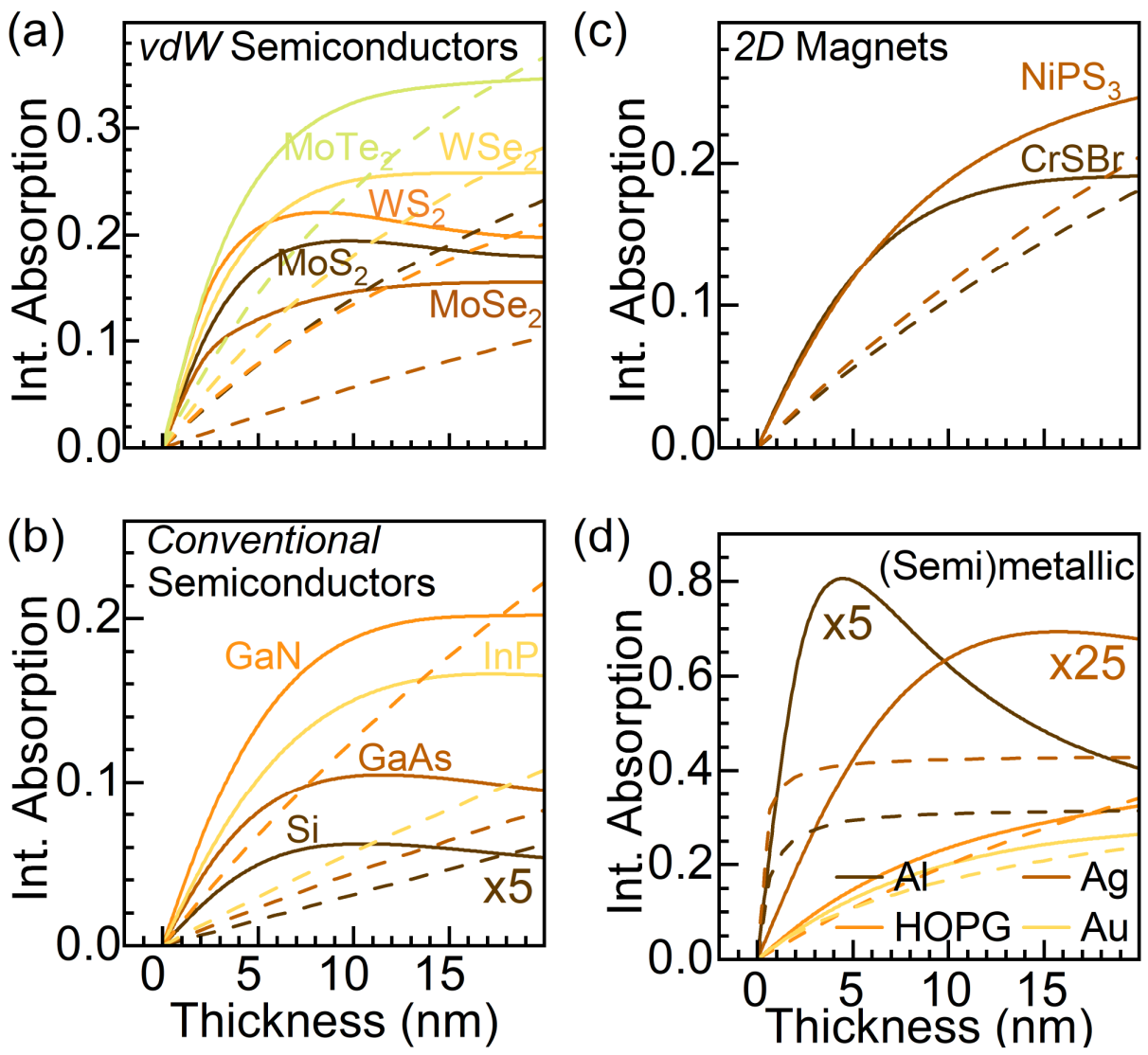

**Figure S7.** This figure is the zoomed in view of Fig. 3 from main text from 0 – 20 nm thickness. Solid and dashed lines show integrated absorption spectra using GTM method and Beer's law for (a) vdW semiconductors; (b) freestanding conventional semiconductors; (c) layered 2DMM on 0.5mm thick sapphire, (d) freestanding metals and c-HOPG (graphite on 0.5mm thick sapphire). The curves for $WS_2$ and $MoSe_2$ are from the present work, other curves are generated using refractive indices obtained from the literature mentioned in Table 1 of main text.

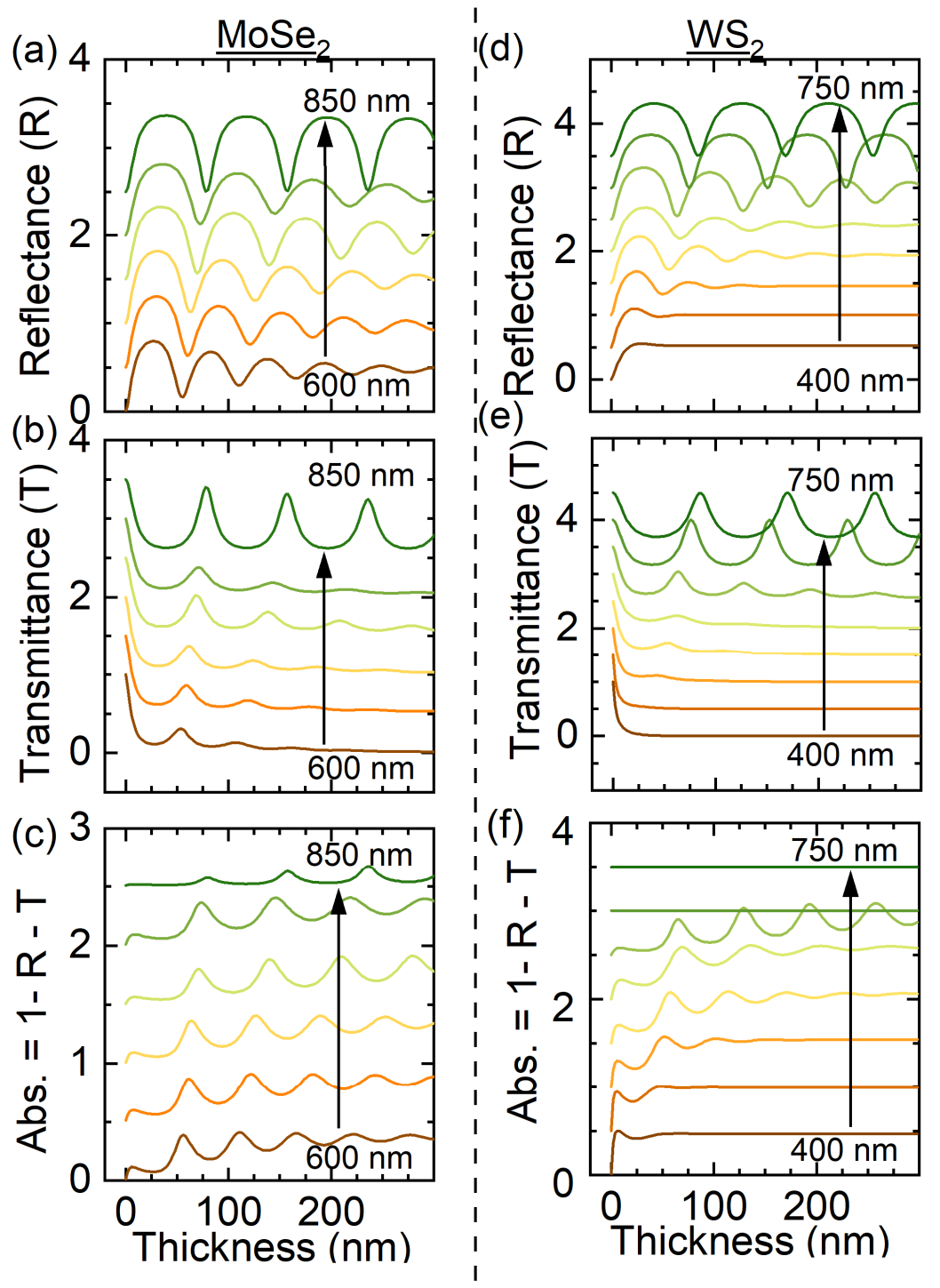

**Figure S8.** R, T and A calculated as a function of layer thickness for freestanding layers (no substrate) of (a–c) $MoSe_2$ and (d-f) $WS_2$ evaluated at selected wavelengths at 50 nm interval. For clarity, successive curves are vertically offset by 0.5 units. The oscillatory features due to interference effects continue to persist in the absorption curves as well. The corresponding integrated absorption curves are given in Fig. S8.

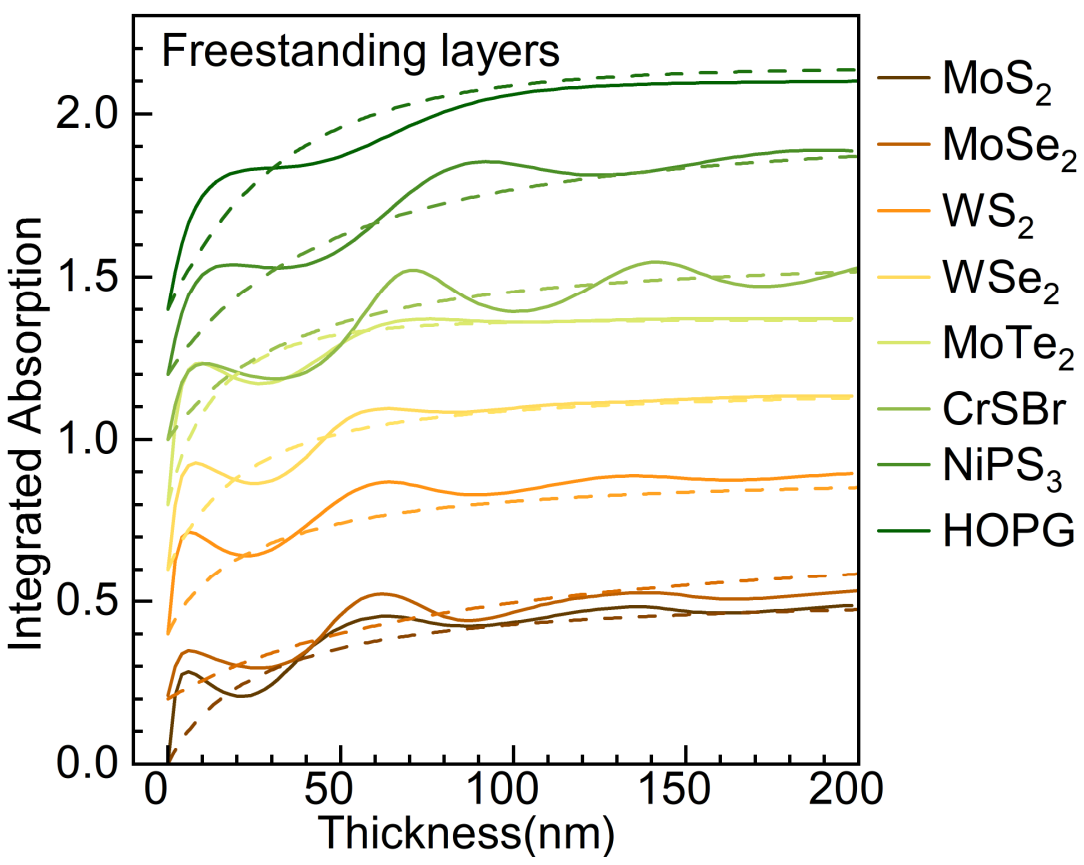

**Figure S9.** Calculated integrated absorption spectra of the freestanding $MoS_2$, $MoSe_2$, $WS_2$, $WSe_2$ and $MoTe_2$, CrSBr, $NiPS_3$, c-HOPG (graphite), using GTMM, as a function of layer thickness with offset of 0.2 on y- axis. Clear oscillatory behavior riding on Beer-Lambert absorption is visible in all the materials.

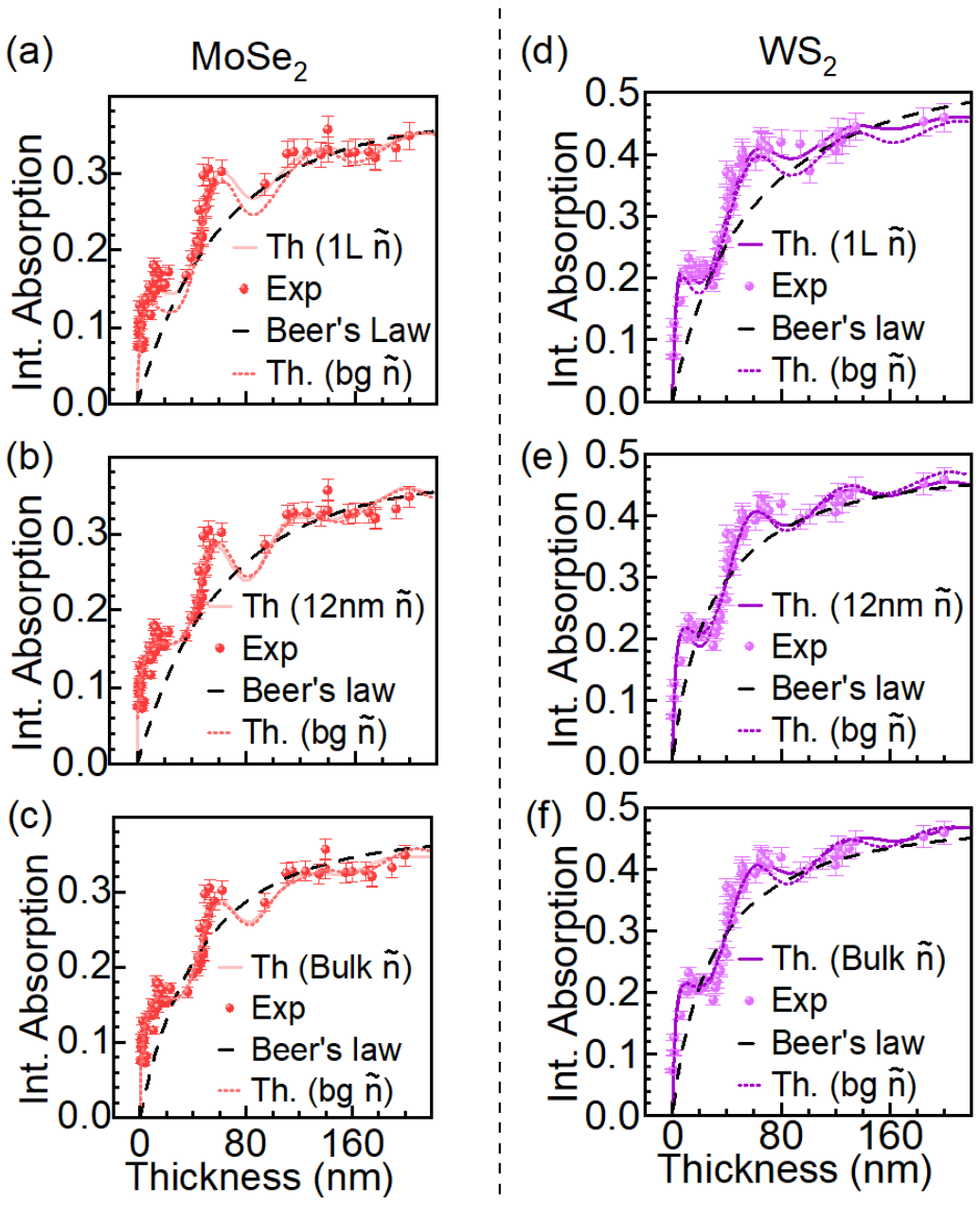

**Figure S10.** Spheres in (a) and (d) are the experimental integrated absorption for $MoSe_2$ and $WS_2$ with varying thickness. Dashed black lines are obtained using Beer's law. Solid lines are the theoretical curves obtained using complex refractive indices ($\tilde{n}$) of monolayer thick crystals (Ref. [9]), while dotted lines are the curves obtained when exciton resonances are removed from $\tilde{n}$. (b) and (e) are similar to (a) and (d), respectively, except that the refractive indices of 12 nm thick films are used. (c) and (f) are similar to (a) and (d), respectively, except that refractive indices of bulk materials are used [9].

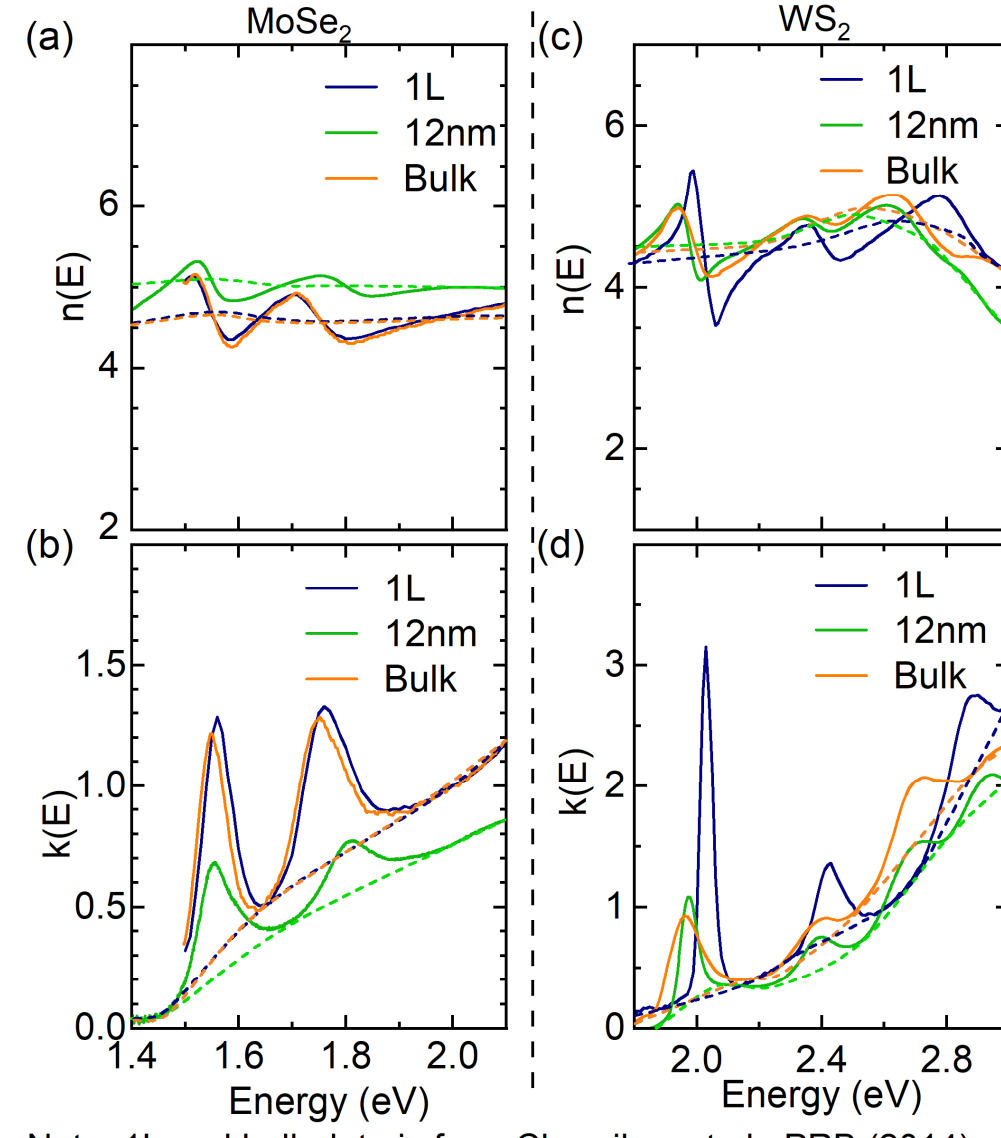

**Figure S11.** Solid lines: Real [n(E)] and imaginary [k(E)] refractive indices of 1L, 12 nm and bulk $MoSe_2$ and $WS_2$ as a function of energy. Dashed lines: refractive indices after removing excitonic contributions. 1L and bulk data is from Ref. [9], while 12 nm curves are obtained in the present work.

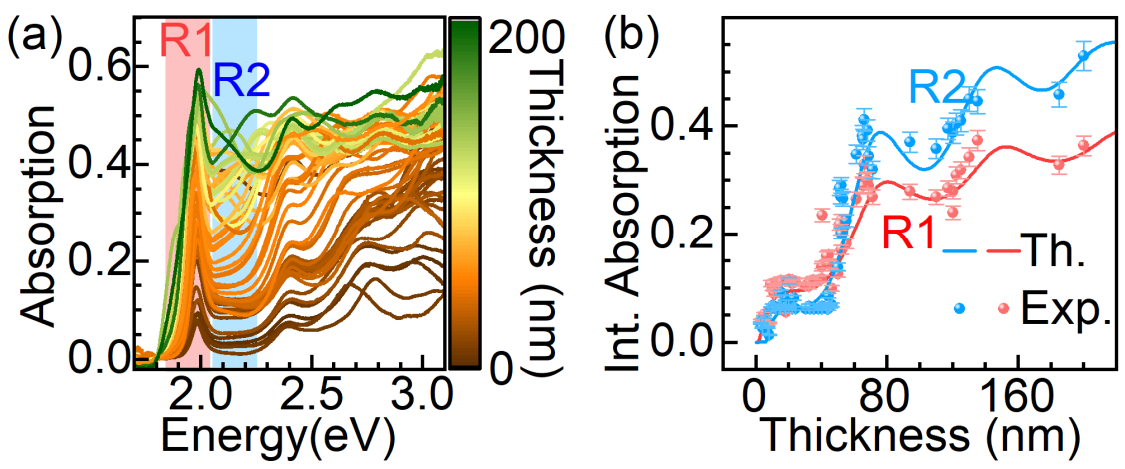

**Figure S12.** (a) Experimental absorption spectra for $WS_2$ with varying thicknesses, ranging from monolayer (1L) to 200 nm. **(**b) Integrated absorption with thickness for two energy windows: one R1 within the excitonic region (1.84-2.04 eV), another R2 away from the exciton resonance (2.05-2.25 eV). The general oscillatory effect in the integrated absorption persists even for the excitonic region or away from the excitonic region.

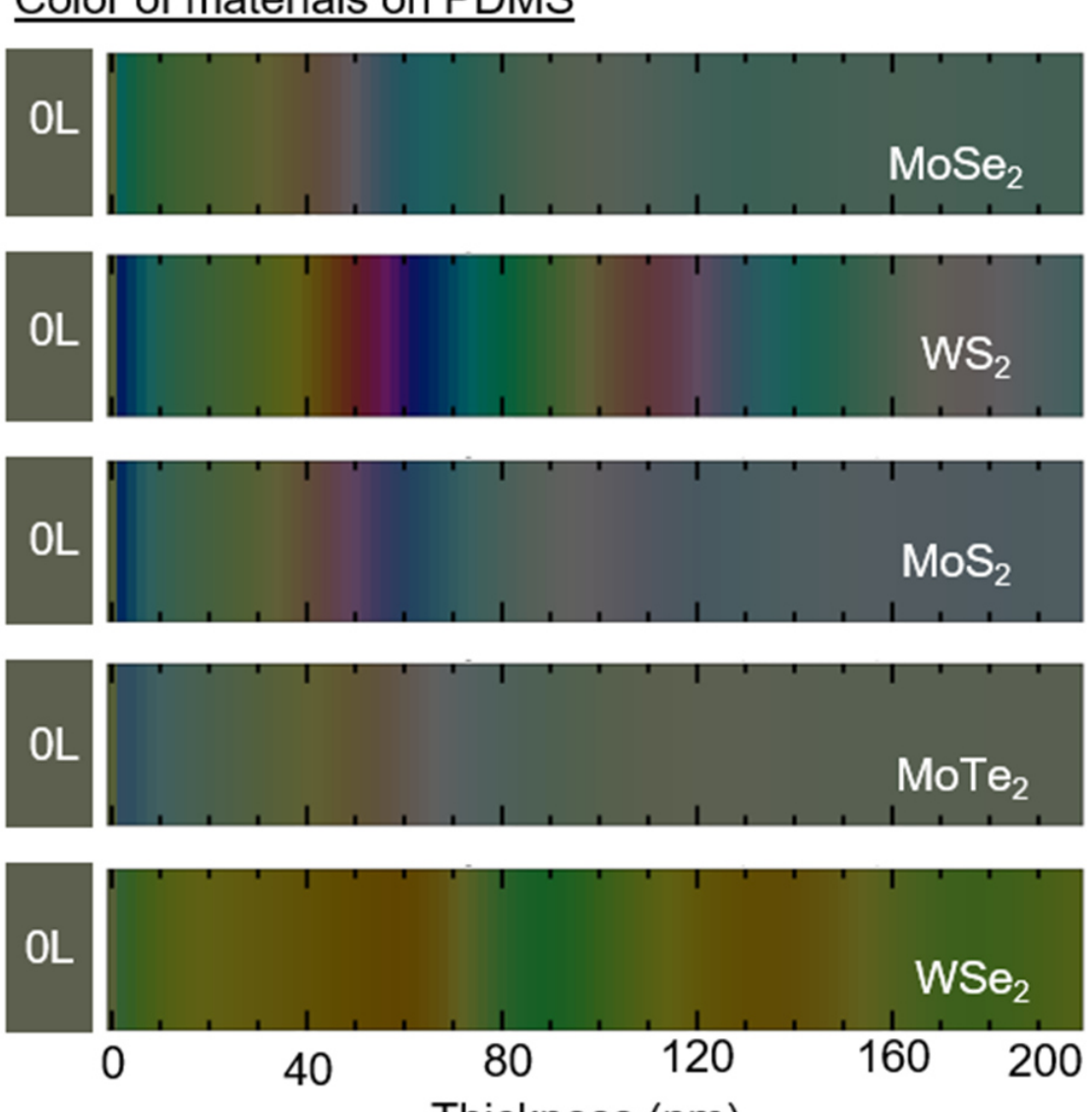

**Figure S13**. Perceived optical colors of $MoSe_2$, $WS_2$, $MoS_2$, $MoTe_2$, and $WSe_2$ on a polydimethylsiloxane (PDMS) as a function of layer thickness, calculated using transfer-matrix method. The leftmost panel ('0L') shows the calculated color of bare PDMS. The calculations are performed assuming illumination with a broadband white-light LED. With increasing thickness, each TMDC exhibits a characteristic progression of interference-induced colors. These distinct color sequences provide a rapid optical means for identifying and distinguishing thickness variations in TMD layers on PDMS.

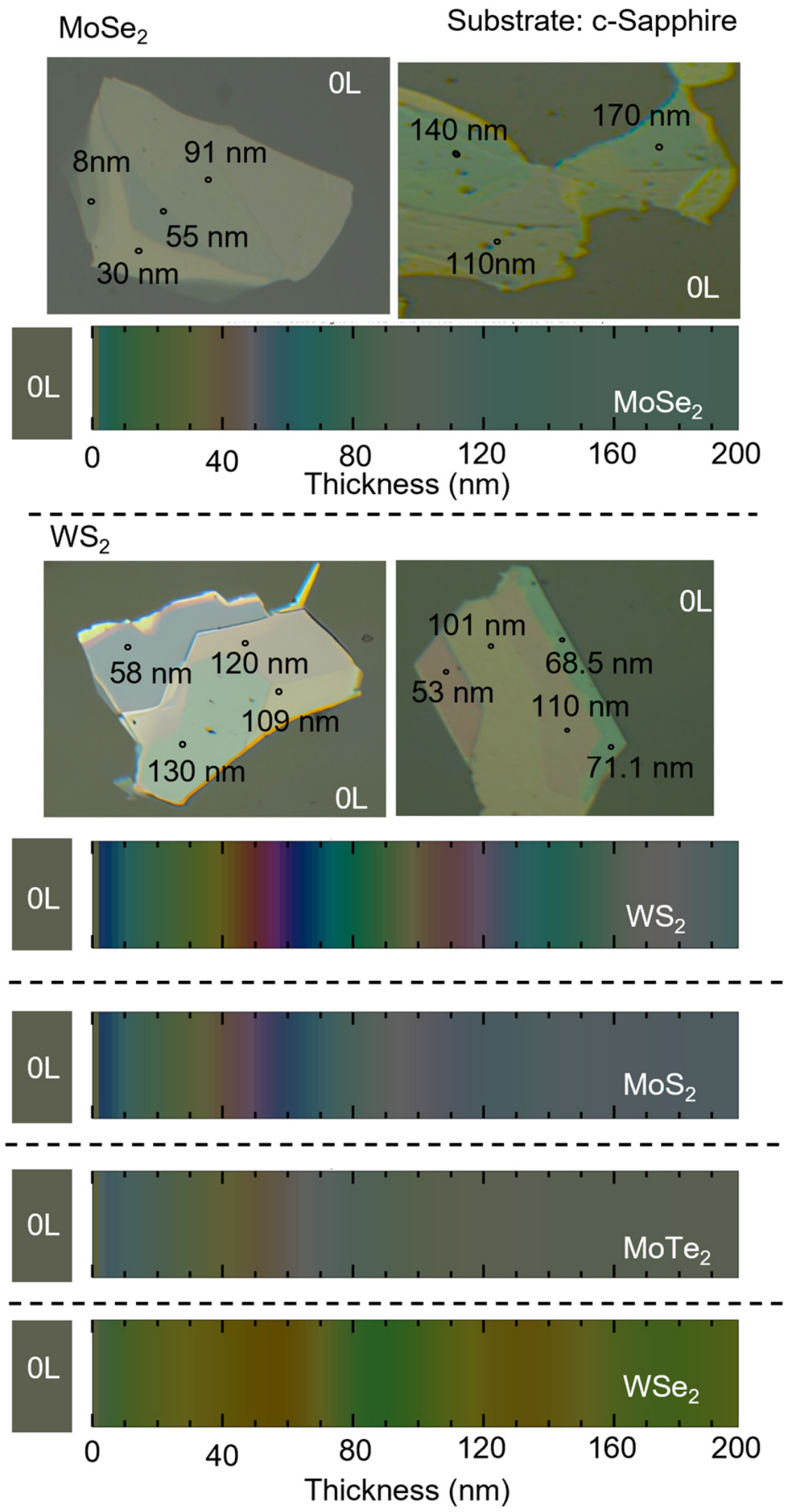

**Figure S14.** Optical images of exfoliated $MoSe_2$ (top) and $WS_2$ (middle) flakes on a 0.5 mm thick sapphire substrate. Thickness values marked on the flakes are obtained from atomic force microscopy (AFM) measurements. At the bottom of the optical images, a color bar is shown, which displays the expected color of the flakes of certain thicknesses kept on c-sapphire calculated using transfer-matrix method. The calculations are performed assuming illumination with a broadband white-light LED. The regions labeled '0L' show the calculated color of bare substrate, providing the reference contrast for identifying the TMDC layers. The colors predicted by the calculations agree excellently with the colors in optical images. Simulated optical color maps of $MoS_2$, $MoTe_2$, and $WSe_2$ on the same substrate are also shown, highlighting the distinct thickness-dependent color evolution across different TMDCs.

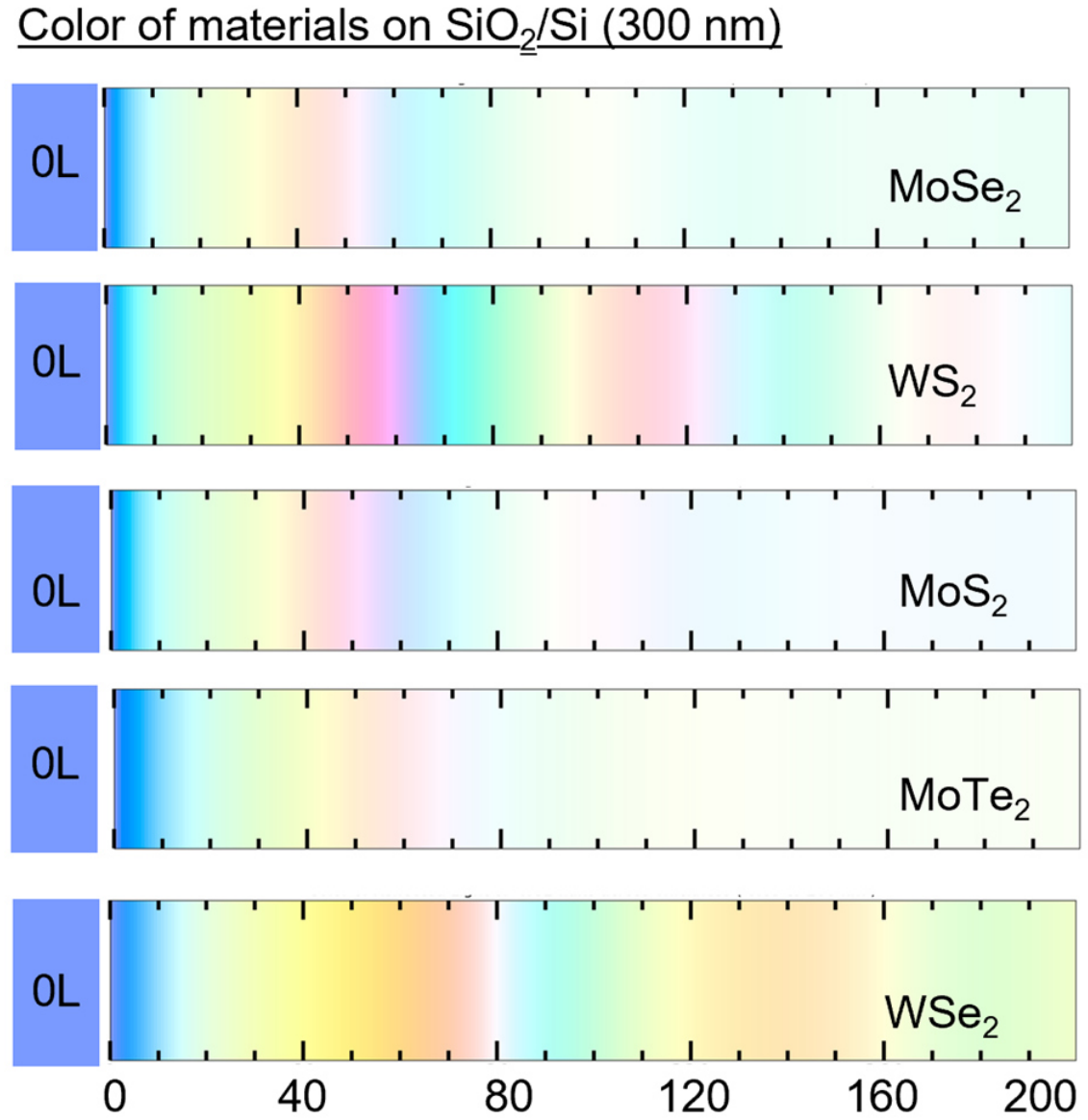

**Figure S15**. Perceived optical colors of $MoSe_2$, $WS_2$, $MoS_2$, $MoTe_2$, and $WSe_2$ on a 300nm $SiO_2$/Si substrate as a function of layer thickness calculated using transfer-matrix method. The leftmost panel ('0L') shows the calculated color of bare 300nm $SiO_2$/Si. The calculations are performed assuming illumination with a broadband white-light LED.

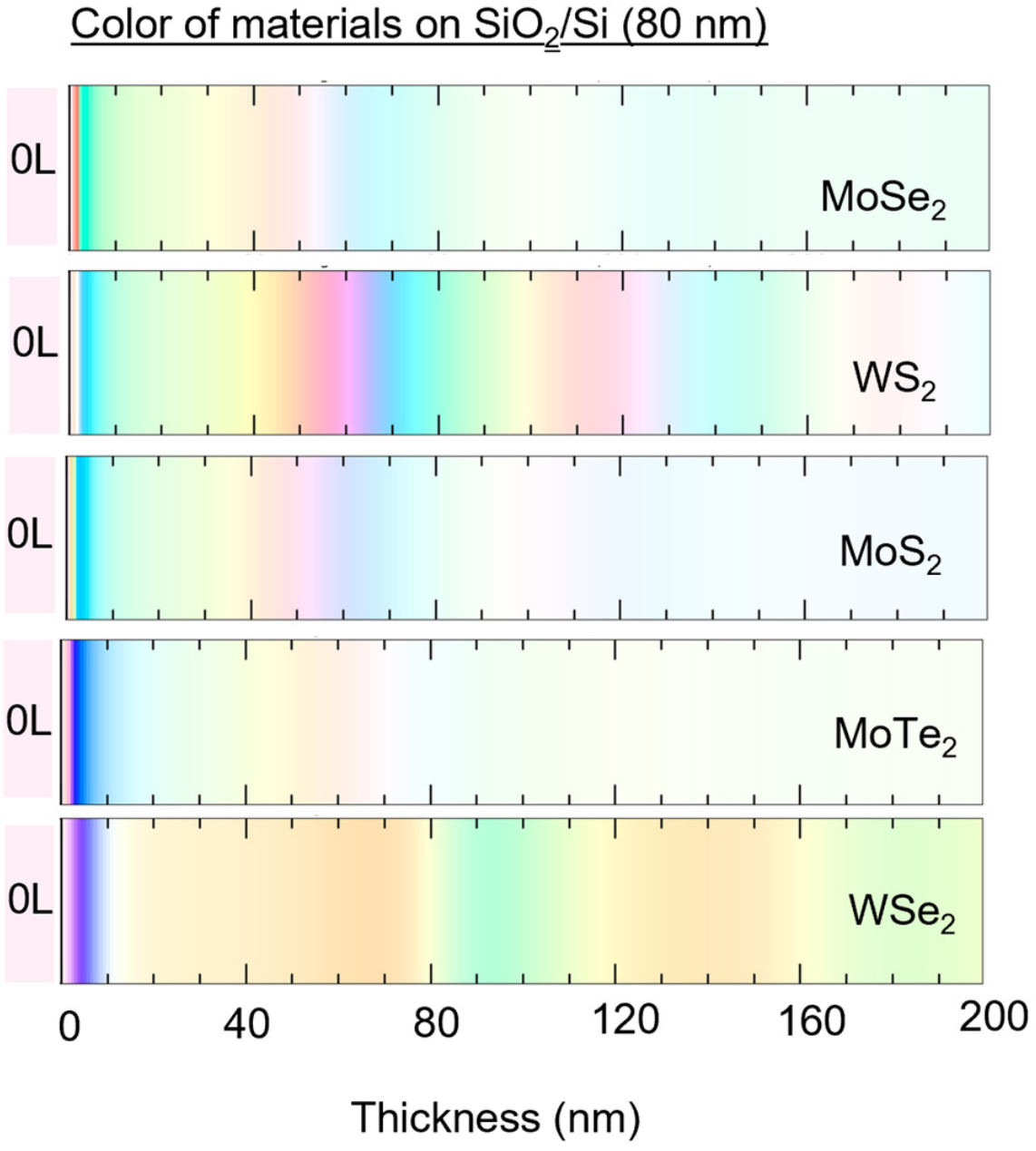

**Figure S16**. Perceived optical colors of $MoSe_2$, $WS_2$, $MoS_2$, $MoTe_2$, and $WSe_2$ on a 80nm ($SiO_2$/Si) substrate as a function of layer thickness calculated using transfer-matrix method. The leftmost panel ('0L') shows the calculated color of bare 80nm $SiO_2$/Si. The calculations are performed assuming illumination with a broadband white-light LED.

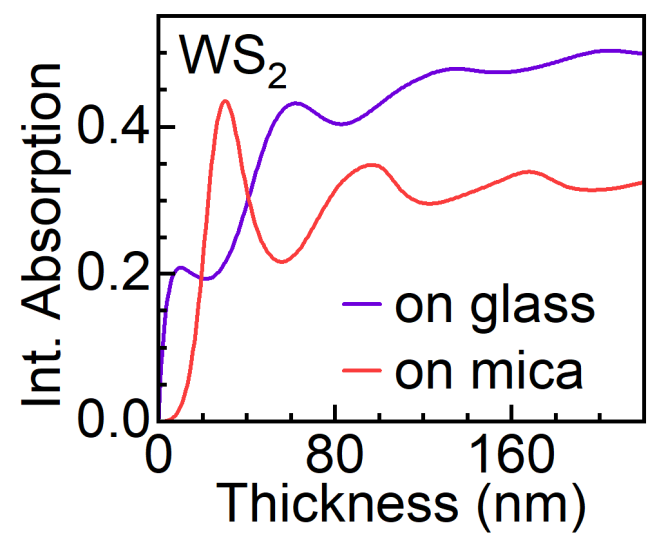

**Figure S17:** Simulated integrated absorption spectra of $WS_2$ as the function of thickness, on glass and mica substrates. The general oscillatory feature of the integrated absorption is independent of the substrate. Energy-dependent refractive indices of glass and mica are taken from Refs. [7] and [8] respectively.

**References.**

[1] M. Fox, *Optical Properties of Solids* (OUP Oxford, 2010).

[2] J. I. Pankove, Optical process in semiconductors, Dover Publ. Inc. **119**, 450 (1975).

[3] D. C. Look and J. H. Leach, On the accurate determination of absorption coefficient from reflectanceand transmittance measurements: Application to Fe-doped GaN, J. Vac. Sci. Technol. B, Nanotechnol. Microelectron. Mater. Process. Meas. Phenom. **34**, (2016).

[4] E. Hecht, *Optics*, 4th ed. (Pearson Addison Wesley, Reading, 2001).

[5] E. Centurioni, Generalized matrix method for calculation of internal light energy flux in mixed coherent and incoherent multilayers, Appl. Opt. **44**, 7532 (2005).

[6] R. Santbergen, A. H. M. Smets, and M. Zeman, Optical model for multilayer structures with coherent, partly coherent and incoherent layers, Opt. Express **21**, A262 (2013).

[7] M. Rubin, Optical properties of soda lime silica glasses, Sol. Energy Mater. **12**, 275 (1985).

[8] R. Nitsche and T. Fritz, Precise determination of the complex optical constant of mica, Appl. Opt. **43**, 3263 (2004).

[9] Y. Li, A. Chernikov, X. Zhang, A. Rigosi, H. M. Hill, A. M. van der Zande, D. A. Chenet, E.-M. Shih, J. Hone, and T. F. Heinz, Measurement of the optical dielectric function of monolayer transition-metal dichalcogenides: $MoS_2$, $MoSe_2$, $WS_2$, and $WSe_2$, Phys. Rev. B **90**, 205422 (2014).